\documentclass[%
 reprint,
 superscriptaddress,
 nofootinbib,
 amsmath,amssymb,
 aps,
 prx,
]{revtex4-2}

\usepackage{graphicx}
\usepackage{dcolumn}
\usepackage{bm}
\usepackage{etoolbox}
\usepackage[colorlinks,linkcolor=blue,urlcolor=blue, citecolor=blue,breaklinks=true]{hyperref}
\usepackage{lipsum}

\usepackage{dsfont}
\usepackage{physics}
\usepackage{enumitem}
\usepackage{tcolorbox}
\usepackage{needspace}
\usepackage[normalem]{ulem}

\usepackage{amsthm} 
\usepackage{thmtools} 
\usepackage{thm-restate}

\declaretheorem{corollary,lemma,proposition,definition}

\usepackage{amsmath, amssymb, amsfonts, amsthm}

\usepackage[page]{appendix}

\renewcommand{\appendixtocname}{List of appendices}

\makeatletter
\let\oldappendix\appendices

\g@addto@macro\tableofcontents{%
  \let\tf@toc@orig\tf@toc
}
\renewcommand{\appendices}{%
  \renewcommand{\thesection}{}
  \let\tf@toc\tf@app
  \addtocontents{app}{\protect\setcounter{tocdepth}{1}}
  \immediate\write\@auxout{%
    \string\let\string\tf@toc\string\tf@app
  }
  \oldappendix
}%

\g@addto@macro\endappendices{%
  \let\tf@toc\tf@toc@orig
  \immediate\write\@auxout{%
    \string\let\string\tf@toc\string\tf@toc@orig
  }%
}  

\renewcommand\tableofcontents{%
    \@starttoc{toc}%
}

\newcommand{\listofappendices}{%
  \begingroup
  \newcommand{\contentsname}{\appendixtocname}
  \let\@oldstarttoc\@starttoc
  \def\@starttoc##1{\@oldstarttoc{app}}
  \tableofcontents
  \endgroup
}

\makeatother

\begin{document}

\preprint{APS/123-QED}


\title{Precision is not limited by the second law of thermodynamics}

\author{Florian Meier}
\email[]{florianmeier256@gmail.com}
\affiliation{Atominstitut, Technische Universit{\"a}t Wien, 1020 Vienna, Austria}

\author{Yuri Minoguchi}
\affiliation{Atominstitut, Technische Universit{\"a}t Wien, 1020 Vienna, Austria}
\affiliation{Institute for Quantum Optics and Quantum Information - IQOQI Vienna, Austrian Academy of Sciences, Boltzmanngasse 3, 1090 Vienna, Austria}

\author{Simon Sundelin}
\affiliation{Department of Microtechnology and Nanoscience, Chalmers University of Technology, 412 96 Gothenburg, Sweden}

\author{Tony Apollaro}
\affiliation{Department of Physics, University of Malta, Msida MSD 2080, Malta}
\author{Paul Erker}
\affiliation{Atominstitut, Technische Universit{\"a}t Wien, 1020 Vienna, Austria}
\affiliation{Institute for Quantum Optics and Quantum Information - IQOQI Vienna, Austrian Academy of Sciences, Boltzmanngasse 3, 1090 Vienna, Austria}

\author{Simone Gasparinetti}
\email{simoneg@chalmers.se} 
\affiliation{Department of Microtechnology and Nanoscience, Chalmers University of Technology, 412 96 Gothenburg, Sweden}

\author{Marcus Huber}
\email{marcus.huber@tuwien.ac.at} 
\affiliation{Atominstitut, Technische Universit{\"a}t Wien, 1020 Vienna, Austria}
\affiliation{Institute for Quantum Optics and Quantum Information - IQOQI Vienna, Austrian Academy of Sciences, Boltzmanngasse 3, 1090 Vienna, Austria}

\date{\today}

\begin{abstract}
Physical devices operating out of equilibrium are inherently affected by thermal fluctuations, limiting their operational precision.
This issue is pronounced at microscopic and especially quantum scales and can only be mitigated by incurring additional entropy dissipation.
Understanding this constraint is crucial for both fundamental physics and technological design.
For instance, clocks are inherently governed by the second law of thermodynamics and need a thermodynamic flux towards equilibrium to measure time, which results in a minimum entropy dissipation per clock tick.
Classical and quantum models and experiments often show a linear relationship between precision and dissipation, but the ultimate bounds on this relationship are unknown.
Our theoretical discovery presents an extensible quantum many-body system that achieves clock precision scaling exponentially with entropy dissipation.
This finding demonstrates that coherent quantum dynamics can surpass the traditional thermodynamic precision limits, potentially guiding the development of future high-precision, low-dissipation quantum devices.

\end{abstract}

\maketitle

\begin{figure*}
    \centering
    \includegraphics[width=\textwidth]{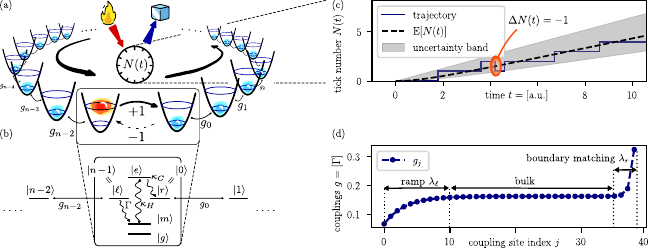}
    \caption{
    The ring clock. (a) Schematic depiction. The clock consists of a ring of $n$ quantum systems (egg cups) hosting a single excitation which travels around the ring.
    Upon completing one cycle, the clock \textit{ticks} by undergoing a biased jump from the last to the first site.
    (b) Level diagram of a quantum system providing a directional interface between the first and last site of the ring using a thermal gradient.
    The level diagram is in the single-excitation subspace, i.e., if one of the sites is in an excited state, all others are in the ground state. See Appendix Sec.~\ref{SM:level_scheme} for the details.
    (c) Representative trajectory of the number of ticks $N(t)$ counted by the clock as a function of time (solid line).
    Due to thermal fluctuations, such a counter can jump backwards (highlighted jump). For the clock to be precise, such backwards jumps must be suppressed, using a strong thermal gradient.
    (d) Numerically optimized couplings, $g_j$, between the nearest-neighbor sites of the ring clock, for a ring of $n=40$ sites. Based on the dependence of the coupling coefficients on the site position in the ring, well approximated by Eq.~\eqref{eq:gn}, we identify three regions. In the initial ramp region of length $\lambda_\ell$, an excitation present in the first site is autonomously shaped into a traveling wave packet. The bulk propagation region is akin to a delay line. Finally, the boundary matching region, of length $\lambda_r$, ensures that the wave packet is absorbed from the last site without reflection.
    }
    \label{fig:visualization}
\end{figure*}

The basic dynamical equations of physics all seem to be invariant under time reversal, i.e.,~symmetric with respect to time.
As systems become more complex, this symmetry is observed to break down.
This, statistical, breaking of time-reversal symmetry through the second law of thermodynamics is fully compatible with reversible microphysics and seems to be the only contender for an explanation for the emergence of a clear notion of past and future in physics.
The implications of this profound insight for the nature of time have long been the center of discussion in the foundation of physics.
Borrowing a dictum from Einstein, \emph{time is what a clock measures}, clocks are the witnesses of the macroscopic breaking of reversibility.
As irreversible out-of-equilibrium systems, clocks come at a fundamental thermodynamic cost -- entropy dissipation~\cite{MilburnReview}.
In the quest for the most accurate clocks, currently based on atomic~\cite{Ludlow2015,Riehle2015,Bothwell2022,Atomic1,Atomic2} or possibly nuclear transitions in the future~\cite{Atomic3}, these costs are not the most pressing concern.
But the quest for small, self-contained quantum control~\cite{Amir2023,Woods2023,Culhane2024,Woods2024,Xuereb2023,Meier2024} shifts the question about the exact relationship between dissipation and precision from a foundational one to a potentially practical one.
The notion of autonomous clocks not requiring external control to run allows us to explore the ultimate dissipation limits of clocks~\cite{Erker2017,Barato2016,Schwarzhans2021,Dost2023,Silva2023}, and may as well inform practical designs for self-contained quantum control~\cite{Meier2024}.

To quantify these limits, one has to resort to microscopic models for the clock.
In such models, all the resources that the clock requires to run are explicitly accounted for within the model and cannot be borrowed from external sources.
Such tiniest conceivable clocks measure time by counting elementary stochastic events as ticks in a regime far away from where state-of-the-art clocks work.
The precision of ticking clocks can be defined as the number of times $\mathcal N$ said clock ticks until it misses one tick compared to parameter time~\cite{Erker2017}, while the corresponding thermodynamic cost is quantified by the entropy $\Sigma_\mathrm{tick}$ dissipated per unit tick.
For fixed $\Sigma_\mathrm{tick}$, one may ask what is the maximum possible clock precision. This question relates the fundamental limit of clock performance to the second law of thermodynamics.
A similar tradeoff concerned with limits on fluctuations of thermodynamic fluxes is encountered in stochastic thermodynamics with the so-called thermodynamic uncertainty relations (TUR).
For classical stochastic systems, these limits are dictated by entropy production.
Classical TUR have received considerable attention~\cite{Barato2015,Horowitz2020}, opening the question whether the same limits apply in the quantum domain~\cite{Liu2019a,Agarwalla2018,Brandner2018,Gerry2022,Prech2023}.
For fully dissipative clocks, a linear bound $\mathcal N\leq \Sigma_\mathrm{tick}/2$ tightly bounds the clock precision.
Such a bound has been confirmed both theoretically~\cite{Erker2017,Barato2016} and experimentally~\cite{Pearson2021}.
In certain quantum scenarios, the linear bound can be beaten by using quantum coherence beyond the dissipative regime.
So far, however, only small theoretical violations have been reported~\cite{Liu2019a,Agarwalla2018,Prech2023}, and larger ones remain contested~\cite{Brandner2018,Gerry2022}.

Here, we report the discovery of a fully autonomous quantum clock model whose precision grows with entropy production as
\begin{align}
\label{eq:N_exp_Sigma}
    \mathcal N = e^{\Omega(\Sigma_\mathrm{tick})}\ ,
\end{align}
exponentially surpassing the dissipative TUR~\cite{Barato2015}.
The $\Omega$-notation denotes an asymptotic lower bound, ignoring constant factors~\cite{Knuth1976}.
The proposed quantum clock is based on a spin chain with site-dependent nearest-neighbor couplings, which could be realized extensibly, for example, in the circuit quantum electrodynamics architecture with coupled cavity arrays~\cite{blais2021,jouanny2024}, and the setup as sketched in Fig.~\ref{fig:visualization}(a,b). 
The clock works by topologically closing the spin chain to a ring and transporting a single excitation around it.
A chirality is introduced by applying a thermal bias between the first and last site.
This setup, which we refer to as the \textit{ring clock}, counts the net number of completed cycles as ticks (see Fig.~\ref{fig:visualization}(c)).

We obtain the exponential scaling by numerically optimizing the coupling coefficients between the sites in the ring.
The key physical principle to enable this scaling is the dissipation-free coherent transport in the bulk of the ring: by adding more sites, the clock's precision can be arbitrarily increased while the dissipation only occurs between the two sites closing the ring, and therefore does not grow with the ring size.
We also provide a quantitative interpretation of the obtained coefficients in terms of wave-packet reshaping and boundary matching, highlighting a connection between our model and previous techniques from optimal coherent quantum transport~\cite{Christandl2004,Apollaro2012}, dissipative quantum transport in condensed matter~\cite{Datta2005a,Fedorov2021}, and photonics~\cite{Sumetsky2003,Chak2006,ferreira2021}.

\paragraph*{Impact.}
Our findings resolve a long-standing foundational question about the ultimate relationship between clocks and the second law of thermodynamics. At the same time,
the ring clock holds promise beyond such fundamental considerations.
In thermodynamic terms, the coherent transport in a degenerate subspace of a spin chain presents a paradigm shift in the way thermal machines on the quantum scale are conceived, paving the way towards extensible quantum machines that exhibit a thermodynamic advantage over their classical counterparts.
From the perspective of quantum control, our results provide a pathway towards high-fidelity and near-dispersion-free transport of quantum information across an array of spins.

\section{\label{sec:results}Theoretical description}

\paragraph*{Model.}
The model we work with is based in the single-excitation regime of a spin chain.
Thus, the relevant basis states can be written as $\ket{0}:=\ket{10\cdots 0}$ for the state where the excitation is on the first site, $\ket{1}:=\ket{01\cdots 0}$ for the state where it is in the second site and so on until $\ket{n-1}:=\ket{0\cdots 01},$ as visualized in Fig.~\ref{fig:visualization}(a).
With coherent nearest-neighbor hopping interactions, we obtain the single-excitation subspace Hamiltonian,
\begin{align}
\label{eq:H}
    H = \sum_{j=0}^{n-2} g_{j}\ketbra{j}{j+1} + \mathrm{h.c.}
\end{align}
The constant $g_{j}$ is a real parameter describing the coherent hopping rate between site $j$ and $j+1$.
In this form, the chain of sites does not yet topologically form the desired ring.
A dissipative coupling between the first site $\ket{0}$ and the last site $\ket{n-1}$ closes the loop and is described by a Lindblad jump operator of the form $J = \sqrt{\Gamma}\ketbra{0}{n-1}$ to model the jump in the direction $\ket{n-1}\rightarrow \ket{0}$.
In Fig.~\ref{fig:visualization}(b) we propose a level-scheme that allows for such a jump process between the last and first ring site (additional details in Appendix~\ref{SM:level_scheme}).

\paragraph*{Entropy production.}
At finite entropy production, local detailed balance predicts that each jump process is accompanied by its time-reverse that is suppressed by how much entropy is produced in each jump.
Here, this is described by the process $\overline J = e^{-\Sigma_\mathrm{tick}/2}J^\dagger$ whose rate is suppressed by the factor $e^{-\Sigma_\mathrm{tick}}$ from detailed balance, and $\Sigma_\mathrm{tick}$ is the entropy produced by each unit population that undergoes the forward transition $\ket{n-1}\rightarrow\ket{0}$.
Given the initial state $\rho(0)=\ketbra{0}{0},$ we describe the evolution of the system using a quantum master equation $\dot \rho = -i[H,\rho] + \mathcal D[J]\rho + \mathcal D[\overline J]\rho$ in units of $\hbar=1$.
Moreover, the dissipator is defined as $\mathcal D[J]\rho=J \rho J^\dagger - \frac{1}{2}\{J^\dagger J,\rho\}$, with anticommutator $\{ A,B\} = A B + BA$.

\paragraph*{Timekeeping figures of merit.}
The jumps generated by $J$ are counted as positive ticks and the reverse $\overline J$ as ``negative ticks'', giving the number of ticks $N(t)$ as the net completed clock cycles to estimate parameter time $t$.
Clock precision can be quantified using the inverse Fano factor~\cite{Silva2023},
\begin{align}
\label{eq:accuracyN(t)}
    \mathcal N_{\Sigma} = \lim_{t\rightarrow\infty}\frac{\mathrm{E}[N(t)]}{\mathrm{Var}[N(t)]},
\end{align}
comparing expectation value $\mathrm{E}[N(t)]$ to the fluctuations $\mathrm{Var}[N(t)]$.
For high precision, fluctuations should ideally be minimal compared to the expected number of ticks.
The majority of ticks must therefore be positive, meaning the forward jump $J$ should dominate over the backwards one $\overline J$, which requires, a priori, a high entropy production per tick $\Sigma_\mathrm{tick}$.

The goal we aim for, however, is the maximization of $\mathcal N_\Sigma$ by varying the Hamiltonian $H$ while at the same time minimizing the entropy production per tick $\Sigma_\mathrm{tick}$.
To solve this problem, we first work in the regime without the negative ticks, $\overline J \rightarrow 0$, requiring divergent entropy production $\Sigma_\mathrm{tick}\rightarrow\infty$, so that only precision has to be maximized without having to handle $\Sigma_\mathrm{tick}$.
As we find out later, the infinite entropy production is not needed to maintain the high precision, and the results actually hold even when the entropy production is negligibly smaller than the clock precision.

Working without negative ticks simplifies the problem of maximizing precision by making it equivalent to the problem of minimizing the relative variance of the waiting time $T$ between ticks.
The reason being that because the clock resets to the same state $\ket{0}$ after every tick, $N(t)$ can be mapped to a renewal process~\cite{Cox1962,Silva2023} and thus, the central limit applies (details in Appendix~\ref{SM:calcualting_precison_reversible}), giving
\begin{align} \label{eq:accuracyT}
    \mathcal N_\Sigma \rightarrow \mathcal N_\infty = \frac{\mathrm{E}[T]^2}{\mathrm{Var}[T]},
\end{align}
where $\mathrm{E}[T]$ is the expected time between two ticks and $\mathrm{Var}[T]$ the variance.
This precision $\mathcal N_\infty$ defined relative to $T$ is what has been traditionally considered in the field of quantum clocks as the main figure of merit~\cite{Erker2017,Woods2019,Woods2022,Dost2023,Silva2023,Meier2023}.

\paragraph*{Numerical optimization.}
Working in the waiting time picture, we can maximize the clock precision by maximizing $\mathcal N_\infty$ as defined in~\eqref{eq:accuracyT}.
Because clock precision is time scale invariant, we can fix without loss of generality the jump rate $\Gamma$, and determine the coupling constants $g_{j}$ of the Hamiltonian $H$ that maximize $\mathcal N_\infty$.
A global numerical maximization yields coupling constants $g_{j}$ that split the ring into three regions as shown in Fig.~\ref{fig:visualization}(d).
Physically, the three regions are responsible for:
\begin{enumerate}[label=(\arabic*)]
    \item Wave-packet preparation ramp with increasing couplings on a length scale $\lambda_\ell$.
    \item Propagation region in the middle of the ring with flat couplings.
    \item Emission region at the end of the ring apodized on a length scale $\lambda_\ell$, to prevent reflection.
\end{enumerate}
We find that the site dependence of the coupling coefficient is well approximated by
\begin{align}
\label{eq:gn}
    g_j = -\mu_\ell e^{-{j}/{\lambda_\ell}} + g + \mu_r e^{({j-(n-1)})/{\lambda_r}},
\end{align}
with $\mu_\ell,$ $g,$ and $\mu_r$ variable coupling parameters and $\lambda_\ell,$ and $\lambda_r$ the length scale of the exponential ramps (numerically optimized parameters in Appendix Fig.~\ref{fig:params_exp_scaling}).

\paragraph*{Theory: (1) Preparation ramp.}
The excitation in the ring clock that is initially localized on $\ket{0}$ is transformed by the initial ramp of couplings into a wider wave packet propagating clockwise along the ring.
In the limit of large values for the number of ring sites $n,$ the wave packet propagation can be described using a continuum description, where we define the real space coordinates as $x_j=j/\lambda_\ell.$
The particle density $n(t,x_j)=\lambda_\ell^{-1}{|\braket{j}{\psi(t)}|^2}$ can be described by the evolution equation $\partial_t n(t,x) = -2\partial_x(g(x)n(t,x))$, where $g(x)$ is defined by the coupling constant $g(x_j) = g_j$.
The equations of motion for $n(t,x)$ follows from the continuum limit of the Schrödinger equation $i\partial_t \ket{\psi(t)} = H\ket{\psi(t)},$ and conserves probabilities $1=\int_0^\infty \mathrm{d}x\, n(t,x)$.
The continuum model is valid so long as we consider times $t<n/(2g)$ where the wave packet has not yet reached the right ramp of the coupling potential~\eqref{eq:gn}, and furthermore $\lambda_\ell,d \gg 1,$ to ensure that $g(x)$ and $n(t,x)$ do not vary quickly on the lattice length scale.
The derivation of the continuum model is discussed in further detail in the Appendix~\ref{SM:continuum_description}, where we also provide the analytical solution for $n(t,x)$.
In the continuum limit, we find that the initial distribution $n(0,x)$ is transported to $x>0$ and broadened by the width $\lambda_\ell$ of the initial ramp.
This leads to the scaling form $|\braket{j}{\psi(t)}|^2\sim \lambda_\ell^{-1} f((j-2gt)/\lambda_\ell)^2$ of the wave packet in the limit of large times and and displacement $x,2gt\gg \lambda_\ell,$ where $f$ is a function independent of $n$ and $\lambda_\ell$\footnote{The notation $\sim$ denotes asymptotic proportionality, which is sometimes also referred to using the $\Theta$-notation~\cite{Knuth1976}.}.

\begin{figure}
    \centering
    \includegraphics[width=\columnwidth]{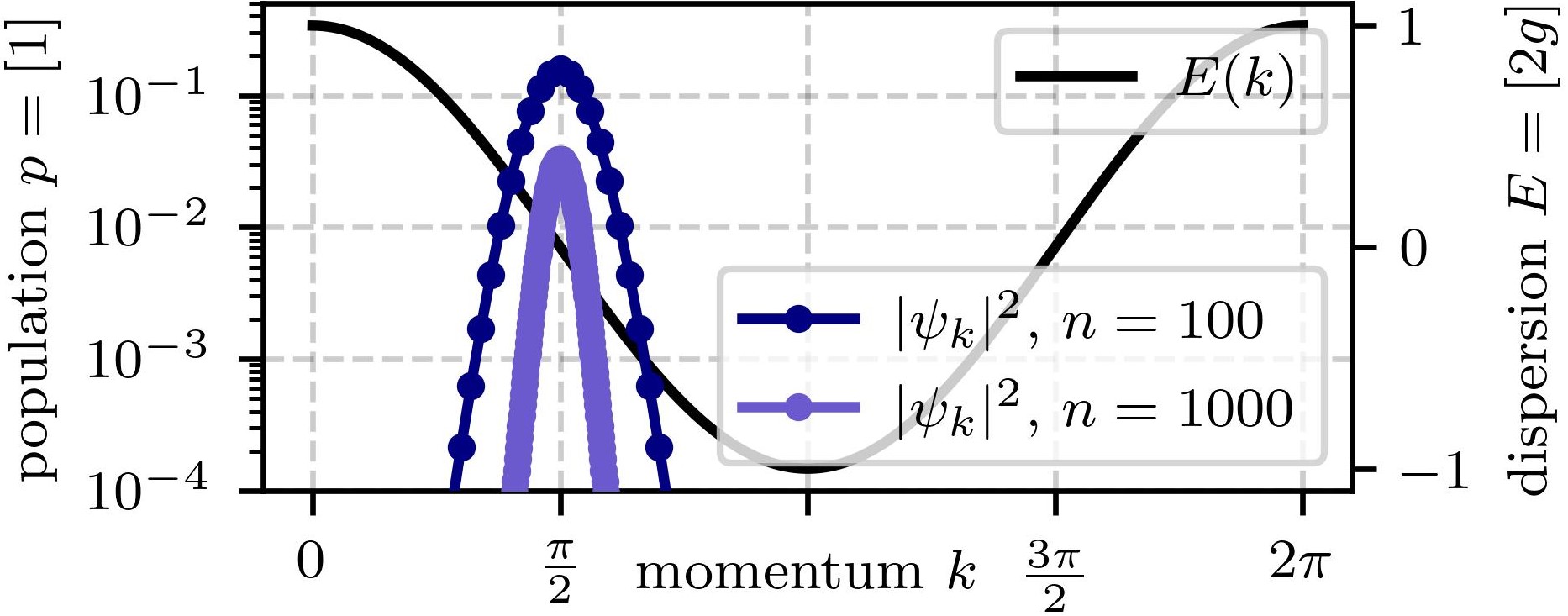}
    \caption{Dispersion relation of the ring, $E(k)$, for a continuum of values $k\in[0,2\pi)$ (solid line, right axis),
    together with a semi-log plot of the momentum space distribution $|\psi_k|^2 = |\braket{\psi_k}{\psi(t)}|^2$ of two wave packets in the bulk of the ring for $n=100$ and $n=1000$ (filled circles, left axis).
    The wave packets are guaranteed to be in the bulk by choosing the time $t$ as half the expected time taken by the wave packet to travel along the ring, $2gt = n/2$.
    The momentum space distribution is centered around $k_0=\pi/2$, indicating a strong concentration of the wave packet around the linear part of the dispersion. For larger values of $n,$ this distribution becomes narrower.}
    \label{fig:RK_space}
\end{figure}

\paragraph*{(2) Wave-packet propagation.}
In the middle region of the ring we approximate the Hamiltonian by using constant hopping parameters $g_{j}=g$ and by imposing periodic boundary conditions.
This approximation is justified for the wave-packet propagating in the middle of the ring because the left tail is exponentially suppressed and the right tail Gaussian suppressed as discussed before.
In this approximation, the model is well-described by a constant coupling, isotropic $XY$ spin chain Hamiltonian~\cite{Lieb1961} $H=\sum_{j=0}^{n-1}  g\ketbra{j}{j+1} + \mathrm{h.c.},$
where we made the cyclic identification $n\equiv 0.$
The Hamiltonian is diagonalized by the discrete Fourier transform $\ket{\psi_k}=\frac{1}{\sqrt{d}}\sum_{j=0}^ {n-1}e^{-ikj}\ket{j}$ of the basis $\ket{j}$ with the momentum $k\in\frac{2\pi}{n}\mathbb Z_n$, allowing us to analytically determine the dispersion relation $E(k) = 2g\cos\left(k\right),$ shown in Fig.~\ref{fig:RK_space}.
Looking at the $k$-space distribution of the wave-function $\ket{\psi(t)}$ while in the middle region of the ring, we find a strong concentration of the probability around momentum $k_0=\pi/2$ as can be seen in Fig.~\ref{fig:RK_space}.
This is no coincidence, as the $\pi/2$-point of the dispersion relation is exactly where $E(k_0+q)=-2gq + O(2gq^3)$ is linear to leading order.
With the wave packet centered around this point, it behaves like a massless quasi-particle being transported dispersion-free to the right with a velocity $2g$.
Since the wave packet is not centered around $\pi/2$ perfectly sharply, the tails in the momentum space distribution will be subject to the non-linearities of the dispersion relation.
We have found in the preceding part (1) that the length scale $\lambda_\ell$ of the initial coupling ramp is what defines the width of the wave packet in real space.
Going to the momentum space, the length scales are inverted and therefore the width of the wave-packet in $k$-space once it has left the initial region is of order $\lambda_\ell^{-1}.$

\paragraph*{(3) Emission.}
Once the wave packet reaches the end of the ring, it is essential that the excitation is emitted with unit probability without being reflected back.
To achieve such an anti-reflection coating, the couplings at the end of the ring must be \textit{apodized} to ensure transmission of a wide range of propagating modes~\cite{Chak2006,Sumetsky2003}.
For long wave guides such apodization schemes have been found to be universal, that is, the coupling parameters between the sites closest to the emitter are independent of the total number of sites~\cite{Sumetsky2003}.
We recover a qualitatively analogous result, where only the last three couplings significantly differ from the bulk values as shown in Fig.~\ref{fig:visualization}(d) and the values of the couplings increase towards the emitter, independently of $n$ (see also Fig.~\ref{fig:params_exp_scaling} in the SM).
With a reflectionless transmission, the ticking statistics are dominated by the properties of the propagating wave packet in the bulk of the ring, which we have analyzed before.

\begin{figure*}[ht]
    \centering
    \includegraphics[width=\textwidth]{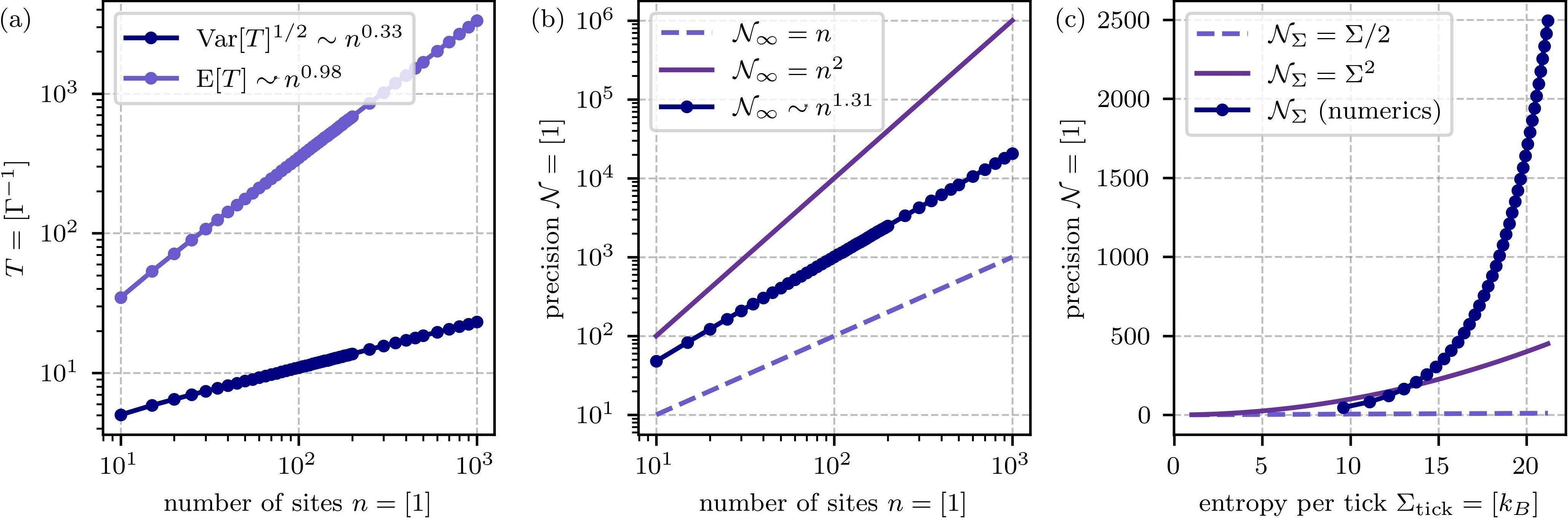}
    \caption{Clock performance \textit{vs.} ring-length, $n$, and entropy production per tick, $\Sigma_\mathrm{tick}$.
    In (a), we show how the expected tick time $\mathrm{E}[T]$ and the standard deviation $\mathrm{Var}[T]^{1/2}$ scale with the number of sites for the numerically optimized choice of couplings.
    The simulation results are in agreement with the prediction that $\mathrm{E}[T] \sim_{\rm th.} n^1$ and $\mathrm{Var}[T]\sim_{\rm th.} n^{2/3}$.
    Numerical values for the exponents are determined by linear regression and the uncertainty in the exponent is of order $10^{-7}$ and thus not shown in the figure.
    Panel (b) shows clock precision $\mathcal N_\infty$ as a function of ring length $n$ in the fully irreversible case (filled circles).
    For comparison, we show the classical and quantum bounds that limit precision by the dimension.
    The precision bound for classical stochastic clocks (dashed line) is linear in the dimension $\mathcal N_\infty \leq n$~\cite{Woods2022} whereas the one for quantum clocks (full line) scales quadratically $\mathcal N_\infty\leq O(n^2)$~\cite{Yang2020,Woods2022}.
    Finally, in (c), we visualize the main result of this work showing that clock precision grows exponentially faster than entropy production $\mathcal N_\Sigma = e^{\Omega(\Sigma_\mathrm{tick})}$ (filled circles).
    We compare this to the TUR bound $\mathcal N_\Sigma \leq \Sigma_\mathrm{tick}/2$ that holds for classical dissipative systems (dashed lines).
    For further comparison, the quadratic scaling $\mathcal N_\Sigma \sim \Sigma_\mathrm{tick}^2$ as recently found~\cite{Dost2023}, ignoring sub-leading terms and constant factors (full line).}
    \label{fig:accuracy_panel}
\end{figure*}

\paragraph*{Tick time statistics.}
The time between two ticks is determined by the tick probability density function (PDF) $p(t)= \Gamma| \langle n-1 \left| e^{-iH_\mathrm{eff}t} \right|0 \rangle |^2$,
with $H_\mathrm{eff}=H-i\frac{1}{2}J^\dagger J$ the effective Hamiltonian of the system including the decay back action from the jump operator $J$ (details in Appendix~\ref{SM:calculating_precision}).
The tick PDF can be split into two contributions $p(t)=p_0(t) + p_1(t)$, where $p_0(t)$ is the free theory without loss term $J^\dagger J$ and without the right ramp, and $p_1(t)$ is the interaction part to restore the equality with the tick PDF.
While the interaction part can formally be obtained as a Dyson series, the free theory is given by the overlap of the wave function with $\ket{n-1}$ at time $t$ (details in Sec.~\ref{SM:emission} of the SM).
To good approximation, $p_0(t)$ dominates the tick time, that is, $\mathrm{E}[T]\sim n/(2g)$ is the time it takes the wave packet to arrive at the final site with the propagation velocity $2g$. 
Similarly, the uncertainty of the arrival time $\mathrm{Var}[T]\sim \lambda_\ell^2 /(2g)^2$ is given by the wave packet's width $\lambda_\ell$, where we use the symbol $\sim$ do denote asymptotic proportionality.

\section{Optimization results}

\paragraph*{Derivation of optimal precision scaling.}
With our previous considerations we find $\mathcal N_\infty \sim n^2 / \lambda_\ell^2,$ and to maximize precision we would naively want to send $\lambda_\ell/n$ to zero.
Because the wave packet's width in momentum space is proportional to $\lambda_\ell^{-1}$, the momentum distribution broadens around $\pi/2$ when $\lambda_\ell\rightarrow 0$.
In this extreme limit, our model of dispersion-free transport breaks down due to higher order skewing effects from the dispersion relation leading to an overall lower precision.
To get the best possible clock precision, we therefore want to have $\lambda_\ell/n$ as small as possible, but not too small such that we avoid running into the issues described above, so how large does it need to be?
For answering this question, the dispersion relation is expanded around $k_0=\pi/2$ to third order, $E(k_0+q)= -2g q + {g q^3}/{3}$, revealing a cubic skewing term, while the linear term is responsible for the propagation.
The scaling $\lambda_\ell \sim n^{1/3}$ guarantees that the cubic term does not scale with $n$ because $gt q^3 \sim n \lambda_\ell^{-3}$ is constant, and thus, the detrimental non-linear effects can be mitigated.
If the exponent were any smaller than $1/3$, the cubic error term would grow as we increase the ring length $n$, and if it were larger, the ratio $\lambda_\ell/n$ would not be at it's theoretical minimum, thus not maximizing clock precision.

\paragraph*{Numerical exact simulation.}
This scaling is confirmed by numerically searching for the values of $g_j$ that maximize $\mathcal N_\infty$.
We find $\lambda_\ell^{\text{num.}} \sim n^{0.35}$, for up to $n=1000$ sites (see Appendix Fig.~\ref{fig:params_exp_scaling}), beyond which computational runtime limited the simulation.
Furthermore the numerics also verify the relationship between $n$ and the tick time statistics, with the scaling laws
$\mathrm{E}[T] \sim n^{0.99}$,
and
$\mathrm{Var}[T] \sim n^{0.66}$, as shown in Fig.~\ref{fig:accuracy_panel}(a).
Combined, we find that clock precision in the theory model and the numerical simulation scales as
\begin{align} \label{eq:Nt_num_vs_thy}
    \mathcal N_\infty 
        \overset{\mathrm{num.}}{\sim}
        n^{1.31} 
        \overset{\mathrm{th.}}{\sim} 
        n^{4/3},
\end{align}
in agreement with each other, see Fig.~\ref{fig:accuracy_panel}(b).

\paragraph*{Precision-entropy scaling.}
We recall that to obtain the precision scaling~\eqref{eq:Nt_num_vs_thy}, we had to assume that the stochastic transition $\ket{n-1}\rightarrow\ket{0}$ was one-way which thermodynamically required divergent entropy production per tick.
By relaxing this idealization and introducing the small perturbative parameter $\delta = e^{-\Sigma_\mathrm{tick}}$, the finite entropy case can be treated analytically.
Negative ticks then enter as a correction of order $\delta$ to the master equation, and the clock precision can be expanded in powers of this correction, $\mathcal N_\Sigma = \mathcal N_\infty + O(\delta)$, using the Landau big-$O$ notation~\cite{Knuth1976}.
Because we aim to minimize entropy $\Sigma_\mathrm{tick}$ we want to keep $\delta$ as large as possible while at the same time making sure the precision $\mathcal N_\Sigma$ is sufficiently close to $\mathcal N_\infty$.

If $\delta = n^{-\beta}$ decays algebraically for some exponent $\beta>0$ large enough to cancel out the constant factors in the big-$O$ correction, the balancing act can be successful.
How large $\beta$ needs to be is determined in part by the spectral gap of the system's Lindbladian and by whether it scales algebraically with system size (which is related to the famously hard problem of undecidability of the existence of spectral gaps~\cite{Cubitt2015}, see Appendix~\ref{SM:precision_entropy_scaling}).
We have numerically determined the scaling up to $n=200$ and found that the choice $\beta = 4$ is sufficient to make the error between $\mathcal N_\Sigma$ and $\mathcal N_\infty$ arbitrarily small as $n$ grows.
Using the identity $\log\delta = -\Sigma_\mathrm{tick}$, we find the entropy scales logarithmically $\Sigma_\mathrm{tick}=\beta \log n$, whereas $\mathcal N_\Sigma \sim n^{1.31}$ scales polynomially (with the negligibable correction), giving the main result
\begin{align*}
\tag{\ref{eq:N_exp_Sigma} rev.}
    \mathcal N_\Sigma = e^{\Omega(\Sigma_\mathrm{tick})},
\end{align*}
that the clock precision is exponentially separated from the entropy production, as also shown in Fig.~\ref{fig:accuracy_panel}(c).

\section{\label{sec:discussion}Outlook}
We have developed a model system with negligible entropic limitations to clock precision and fully compatible with autonomous quantum evolution.
The autonomy is guaranteed by combining non-equilibrium dissipative dynamics with coherent quantum evolution in a spin chain.
Contrarily to other proposals of autonomous quantum clocks which increase their precision by increasing the maximum energy in the system~\cite{Erker2017,Woods2019,Pearson2021,Schwarzhans2021,Dost2023} and thereby also increase the entropy production, the ring clock works in the energy-degenerate single-excitation subspace and through careful design of the interactions manages to exponentially improve the precision versus entropy scaling compared to the previous results.

The idealized setting of our analysis raises the question about stability: how do imperfections impact the performance?
So far, we have assumed the absence of dissipation in the channel along the entire chain, except for the last site.
In realistic settings, each site would have some finite life-time, and so the excitation could also be dissipated along the way leading to a premature tick, decreasing $\mathrm{E}[T]$ and negatively impacting $\mathrm{Var}[T]$.
In essence, any finite dissipation along the sites would lead to a maximum length in which the exponential scaling can still be upheld, before breaking down again.
Akin to fault-tolerance in unitary quantum devices, this limitation for the ring clock is only of technical nature and there is no principal fundamental bound as to how fault-tolerant the ring transport can be other than the fact that all experiments and devices are imperfect to some degree.

Despite these technical challenges, coupled cavity arrays (CCA) for example, have been employed to realize microwave metamaterials with tailored band structures~\cite{carusotto_photonic_2020, castillo-moreno2024, ferreira2021}, also serving as slow-light waveguides~\cite{ferreira2024}.
They consist of an array of capacitively coupled lumped-element superconducting microwave resonators, which can exhibit low intrinsic dissipation (internal quality factors of $Q_i \sim 10^5$), supporting arrays of up to $n=100$ individual sites~\cite{jouanny2024}.
The couplings between adjacent sites are determined by the capacitive network of the circuit and can be engineered to satisfy the prescription in Fig.~\ref{fig:visualization}(d).
The tick generating element of the clock (Fig.~\ref{fig:visualization}(b)) could be achieved using a superconducting artificial molecule comprising one or more artificial atoms.
Through careful design of the band structure of the CCA and transition frequencies of the energy levels, selective emission from certain transitions can be achieved~\cite{ferreira2024} ensuring that only a single excitation is propagating through the CCA.
Further dissipation into the environment for other transitions can be made possible through tailored dissipation engineering between the artificial atoms and microwave waveguides~\cite{murch2012a, aamir2022a} to model the respective thermal baths~\cite{sundelin2024b}.
To detect the ticks of the clock, we envisage two methods.
One method uses continuous, dispersive readout of a particular eigenstate of the molecule, marking a tick by the detection of a quantum jump~\cite{vijay2011, he2022a}.
Alternatively, ticks can be registered by capturing emitted photons with a microwave photodetector~\cite{inomata2016, besse2018, kono2018, lescanne2020a}, although this approach is currently constrained by photon-detection fidelity.

\acknowledgments
The authors thank Ralph Silva, Nuriya Nurgalieva and Yoshihiko Hasegawa for disussions.
MH, PE and YM acknowledge funding from the European Research Council (Consolidator grant `Cocoquest’ 101043705).
YM acknowledges funding by Grant No.~62179 of the John Templeton Foundation.
This project is co-funded by the European Union (Quantum Flagship project ASPECTS, Grant Agreement No. 101080167). SG acknowledges financial support from the European Research Council via Grant No.~101041744 ESQuAT. Views and opinions expressed are however those of the authors only and do not necessarily reflect those of the European Union, REA or UKRI. Neither the European Union nor UKRI can be held responsible for them.

\section*{Source code and data}
All optimal parameters, data and code are available at the public repository~\url{https://github.com/aspects-quantum/ring-clock}.

%


\clearpage\newpage

\section*{Appendices}
\appendix
\listofappendices

\begin{appendices}
\section{Applicability of the master equation}\label{SM:level_scheme}
Using spin chains or coupled cavity arrays to model  the ring clock from the maintext leads to an effective description with the Hamiltonian $H$ as in eq.~\eqref{eq:H} and jump operators $J,\overline J$.
Working with a spin chain, the local Hamiltonians are given by $\frac{\omega}{2}\sigma_{z,j}$ for site $j$ and the nearest neighbor coupling can be modelled by the particle number conserving hopping term $g_j \sigma_{-,j}\sigma_{+,j+1}+\mathrm{h.c.}$
The challenging part for a microscopic description comes from the fact that we have to prevent multiple excitations entering the ring at once to ensure that the description we have used so far is valid.
One way to solve this problem is to treat the first and the last site of the ring clock separately as a single system -- the ticking element -- as proposed in Fig.~\ref{fig:visualization}(b).
To ensure that we recover the effective dynamics described in the main text, this ticking element has to (1) remember when it has emitted an excitation into the ring such that it does not emit another excitation into the ring before it has ticked, and (2) it has to be boundary-matched to the ring couplings $g_j$ to avoid the reflection of the incoming wave packet.

The proposed level-scheme comprises of 5 states and undergoes the following cycle for a tick $\ket{g}\rightarrow\ket{\ell}\rightarrow\ket{m}\rightarrow\ket{e}\rightarrow\ket{r}\rightarrow\ket{g}$.
Note, we identify $\ket{\ell}\equiv \ket{n-1}$ as formally the last ring site in terms of the states of the reduced model, and $\ket{r}\equiv \ket{0}$ as the first site.
The idea of this scheme is that by adiabatically eliminating the intermediate states $\ket{m}$ and $\ket{e}$ from the tick cycle, we recover the jump process $\ket{n-1}\equiv \ket{\ell}\rightarrow\ket{0}\equiv\ket{r}$ generated by $J$ (and the inverse generated by $\overline J$).
To get there, we look into each step of the tick cycle separately: 
When the excitation is in the ring, the ticking element is in the ground state $\ket{g}$.
When the excitation arrives at the site $\ket{n-2}$ in the ring, it couples to the transition $\ket{g}\rightarrow\ket{\ell}$ with strength $g_{n-2}$.
In turn, the state $\ket{\ell}$ decays dissipatively as $\ket{\ell}\rightarrow\ket{n}$ with rate $\Gamma$, which defines the tick.
This is the rate that crucially has to be boundary matched to the couplings $g_j$, and the strength of the reverse process needs to be suppressed with the entropy production $\Sigma_\mathrm{tick}$ by coupling this transition to a cold enough bath.
After this decay, however, the system is not in resonance with the frequency $\omega$ of the ring structure anymore and it has to brought to a higher energy again, for which two additional dissipative drives can be used.
With an additional hot bath driving $\ket{m}\leftrightarrow \ket{e}$ at a rate $\kappa_H$ and an additional cold bath mediating $\ket{e}\rightarrow\ket{r}$ at a rate $\kappa_C$ we can create a population inversion to ensure that, in the end, the state that arrived on the ring ends up on $\ket{r}$.
Then, the ticking element is again on resonance with $\omega$ and the excitation can be coherently coupled to the next site with strength $g_0$, completing the cycle.

When the excitation is in the ring, the ticking element latches to the ground state $\ket{g}$, which is not addressed by the thermal drives, and therefore, no second excitation is emitted into the ring.
So long as $\omega \gg \kappa_H, \kappa_C, \Gamma \gtrsim g_j$ is the hierarchy of energy scales in the problem, the Lindblad master equation is applicable in a local picture with weak interactions despite the many-body nature~\cite{Hofer2017,Shiraishi2024}, and the notion of entropy $\Sigma_\mathrm{tick}$ we used coincides with the thermodynamic entropy `$\beta Q$' from Clausius' law~\cite{Landi2021}.

\section{Details on the numerical optimization of the precision}\label{sec:clock_precision_first_passage_times}
\begin{figure*}
    \centering
    \includegraphics[width=\textwidth]{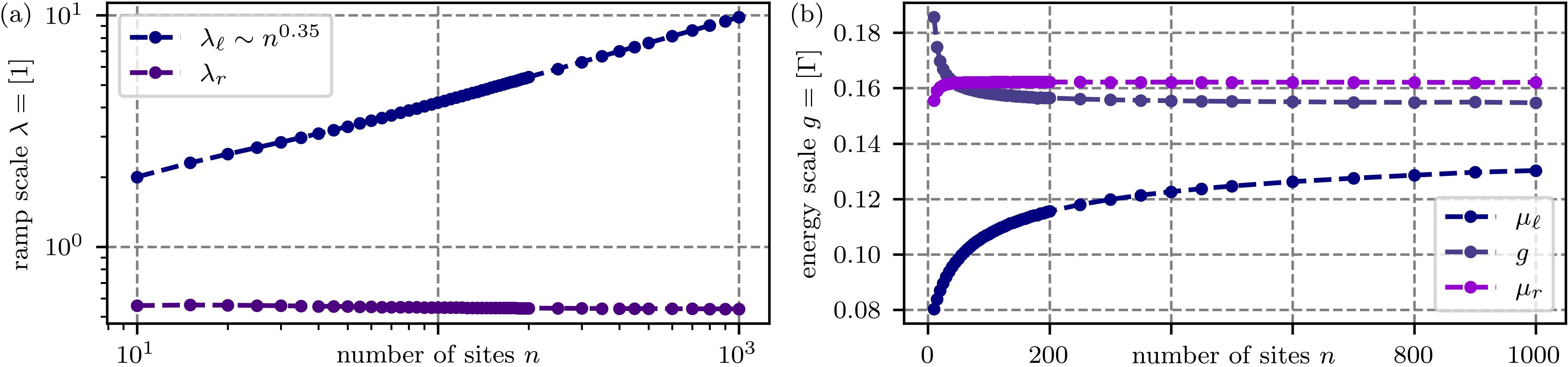}
    \caption{Plot of the numerically optimized parameters $\mu_\ell,\mu_r,g,\lambda_\ell,\lambda_r$ as a function of the ring length $n$.
    In~(a), we show the ramp lengths $\lambda_\ell$ and $\lambda_r,$ finding that the left ramp grows as $n^{0.35}$ close to the predicted exponent of $1/3.$
    The exponent is determined with linear regression and has uncertainty of order $10^{-6}$ and is thus not shown in the figure.
    The right ramp does not become larger with the ring length, indicating an agreement with previous literature on apodization showing that only a constant number of couplings have to modified in a coupled cavity array to avoid reflection at a dissipative sink~\cite{Sumetsky2003}.
    In~(b), we show how the optimal coupling constants $\mu_\ell,\mu_r$ and $g$ change as a function of $n$ showing that both $g$ and $\mu_r$ quickly approach a constant value with growing $n,$ whereas $\mu_\ell$ only slowly increases, indicating that the depth of the left ramp keeps increasing with $n$.}
    \label{fig:params_exp_scaling}
\end{figure*}

In the following, we provide a more in-depth derivation of how it is possible to calculate $\mathcal N_\infty$ given the dynamical description of the clock, that is the Hamiltonian $H$ and the tick generating operator $J$ (Sec.~\ref{SM:calculating_precision}), and how the precision is maximized (Sec.~\ref{SM:maximizing_clock_precision}).
As for how $\mathcal N_\Sigma$ and $\mathcal N_\infty$ are related, we will answer this question in detail in Sec.~\ref{SM:precision_entropy_scaling}.

\subsection{\label{SM:calculating_precision} Clock precision}
Here, we work in the fully irreversible regime without the reverse jump process $\overline J$, and clock precision in this limit can be calculated with the waiting time statistics
\begin{align}
    \mathcal N_\infty = \frac{\mathrm{E}[T]^2}{\mathrm{Var}[T]}.
\end{align}
This quantity has thus been the standard measure for clock precision in the literature~\cite{Erker2017,Woods2022,Dost2023,Meier2023}, and can be interpreted as the number of times the clock ticks on average until it goes off by one compared to a perfect reference clock~\cite{Erker2017}.

We now show how to calculate $\mathrm E[T]$ and $\mathrm{Var}[T]$.
Without reverse ticks, the equations of motion are defined by the Liouvillian $\mathcal L = \mathcal L_0 + \mathcal L_+,$ where the term $\mathcal L_0\,\cdot\, = -i\left[H,\,\cdot\,\right] -\frac{1}{2}\left\{J^\dagger J,\,\cdot\,\right\}$ is the conditional non-trace-preserving part of the evolution where no tick occurs, and $\mathcal L_+\,\cdot\, = J\,\cdot\, J^\dagger$ is the part generating the ticks.
Given some initial state of the clock, $\rho_0,$ the non-normalized evolution given that the clock has not yet ticked is generated by $\mathcal L_0,$ and, by taking the trace, we can find the survival probability $P[T\geq t]$ that at time $t$ the tick has not yet occurred~\cite{Daley2014,Landi2023},
\begin{align}
\label{eq:PTgeqt}
    P[T\geq t] = \tr\left[e^{\mathcal L_0 t}\rho_0\right].
\end{align}
Under the assumption that the resulting probability density function (PDF) $p(t)=P[T=t]$ is normalized, that is $\lim_{t\rightarrow\infty} P[T\geq t]=0,$ we can calculate all moments of $T$ and thereby also the desired quantities $\mathrm{E}[T]$ as well as $\mathrm{Var}[T]$ needed determine the clock precision $\mathcal N_\infty$.
Upon integration over $t\in \mathbb R_+,$ we get
\begin{align}
    \mathrm{E}[T^k] 
    &= 
    k \int_0^\infty \mathrm{d}t\, t^{k-1} \tr\left[e^{\mathcal L_0 t}\rho_0\right]\\
    &= 
    (-1)^k k!\tr\left[\left(\mathcal L_0\right)^{-k}\rho_0\right],
\end{align}
where the expression $\rho_k = \left(\mathcal L_0\right)^{-k}\rho_0$ can be obtained by recursively solving a Lyapunov equation~\cite{Dost2023}. 
To be explicit, the equation $\rho_{k} = \left(\mathcal L_0\right)^{-1} \rho_{k-1}$ is equivalent to the continuous-time Lyapunov equation
\begin{align}
    \label{eq:rho_i_lyapunov}
    \underbrace{\left(-i H + \frac{1}{2}J^\dagger J\right)}_{=:-iH_\mathrm{eff}} \rho_{k} + \rho_{k}\left(-i H + \frac{1}{2}J^\dagger J\right)^\dagger = \rho_{k-1}.
\end{align}
This type of equation arises for when considering stability in quantum master equations~\cite{Rouchon2013,Nicacio2016} and classical linear control theory \cite{Brunton_19}.
Here, we can also identify the non-Hermitian effective Hamiltonian $H_\mathrm{eff}$ as already in the main text.
To calculate the precision fraction $\mathrm{E}[T]^2/\mathrm{Var}[T]$, only the first two iterations are required, and in shortened notation, we can write
\begin{align}
\label{eq:N_T_lindbladians}
    \mathcal N_\infty = \left(\frac{2\tr\left[(\mathcal L_0)^{-2}\rho_0\right]}{\tr\left[(\mathcal L_0)^{-1}\rho_0\right]^2}-1\right)^{-1}.
\end{align}
For our considerations, we use the initial state $\rho_0=\ketbra{0}{0}$, which is the same state the ring clock resets to after a tick $\mathcal L_+ (\rho) \propto \ketbra{0}{0}$ regardless of the state $\rho$ before the tick.

When considering the survival probaility $P[T\geq t]$, it is advantageous to use the purity of the initial (reset) state so the evolution decomposes $e^{\mathcal{L}_0t} \ketbra{0}{0} = e^{-iH_{\mathrm{eff}}t} \ketbra{0}{0} e^{i H_{\mathrm{eff}}^\dagger t}$ which requires the exponentiation of a matrix only of size $O(n^2)$ instead of $O(n^4)$.
Then, the tick probability density is given by
\begin{align}
\label{eq:PT=t}
    P[T=t] 
    &= 
    -\frac{\mathrm{d}}{
    \mathrm{d}t}
    \underbrace{\tr\left[ e^{\mathcal L_0 t}\ketbra{0}{0}\right] }_{ = P[T\geq t]}\\
    &= 
    \Gamma \left| \left\langle n-1 \left| e^{-iH_{\mathrm{eff}}T} \right|0 \right\rangle \right|^2.
\end{align}

\subsection{\label{SM:maximizing_clock_precision} Maximizing clock precision for infinite entropy production}
\begin{figure*}
    \centering
    \includegraphics[width=\textwidth]{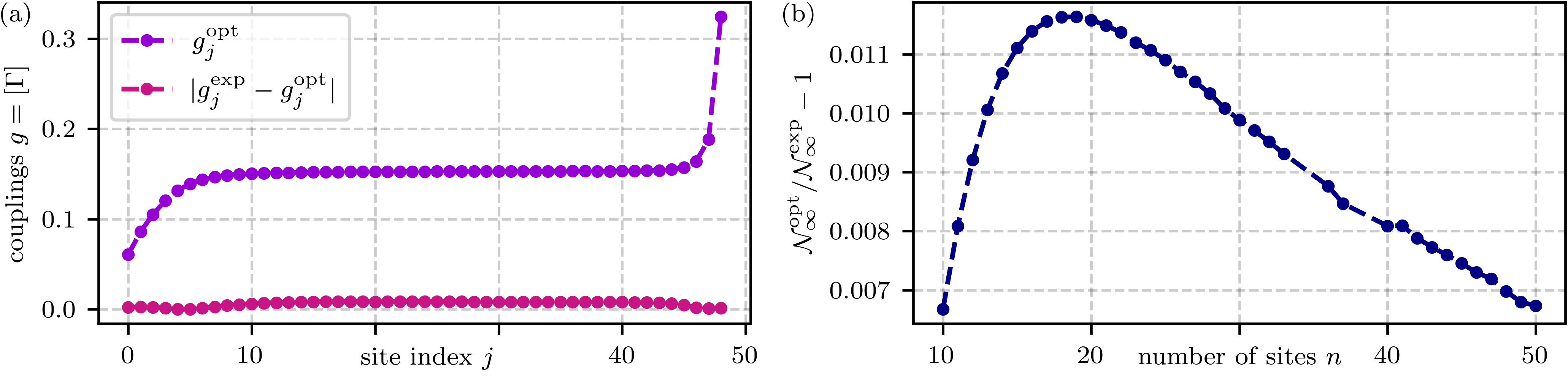}
    \caption{In this panel we show the comparison between the true global optimum of coupling parameters $g_j$ and the optimum for the case where the coupling parameters are given by~\eqref{eq:gn_SM}.
    In (a), we show the difference between the two coupling parameters for the examplary case of $n=50$ ring sites. The difference between the optimal couplings and the ones obtained using the exponential model is at most of order $10^{-2}$ in the bulk and on the boundaries the difference becomes even smaller, i.e., of the order of $10^{-3}.$
    In (b), we plot how the clock precision $\mathcal N_\infty$ differs for those two models. We see that the relative difference vanishes for large site numbers.
    Several outliers due to the numerical optimization have been removed for the site numbers between $n=30$ and $n=40.$
    A comparison becomes unfeasible for higher site numbers due to computational constraints in finding the global optimum with more than $50$ parameters.
    In comparison, the exponential model requires the optimization of only $5$ parameters regardless of the ring length.}
    \label{fig:comparison_true_exp}
\end{figure*}

The clock precision $\mathcal N_\infty$ for fixed initial state $\rho_0=\ketbra{0}{0}$ is a function of the parameters $g_j$ in the Hamiltonian $H = H[\{g_j\}]$ and the ticking rate $\Gamma$ in the jump operator $J$.
Precision is dimensionless $[\mathcal{N}_\infty] = 1$, and thus $\mathcal{N}_\infty$ is invariant under an arbitrary re-scaling of all the rates.
To simplify the analysis, we may therefore fix the parameter $\Gamma$ to some value for the optimization and the remaining parameters $\{g_j\}$ are varied in units of $\Gamma$.
For the most general case we would of course like to determine the global optimum, maximizing over all possible couplings $\{g_j\}_{j=0}^{n-2}$ on the ring,
\begin{align}
    \{g_j\} = \underset{g_j}{\mathrm{argmax}}\big\{\mathcal N_\infty\big\}.
\end{align}
This global optimization of the couplings, however becomes prohibitive $O(n)$ for large system sizes,
while for small values of $n\lesssim 50$ the numerical optimization is stable and converges to the global minimum.
From this solution we extract the ansatz for $g_j$ displayed in Eq.~\eqref{eq:gn} of the main text or
\begin{align}
\label{eq:gn_SM}
    g_j = -\mu_\ell e^{-{j}/{\lambda_\ell}} + g + \mu_r e^{({j-(n-1)})/{\lambda_r}},
\end{align}
which we repeat here for the reader's convenience.
For system sizes beyond $n>50$, we used the ansatz in eq.~\eqref{eq:gn}.
In this case, it is not guaranteed that we find the global maximum of $\mathcal N_\infty$, but since the ansatz in eq.~\eqref{eq:gn_SM} does not include the ring size $n$ any more, the optimization of
\begin{align} \label{eq:opt_5}
    (\mu_\ell,\mu_r,g,\lambda_\ell,\lambda_r) = \underset{\{\mu_\ell,\mu_r,g,\lambda_\ell,\lambda_r\}}{\mathrm{argmax}}\big\{\mathcal N_\infty\big\},
\end{align}
is greatly simplified because only 5 parameters need to be optimized.
Our numerical simulation shows for values $n\lesssim 50$ where both optimization algorithms converge, that the two methods yield qualitatively similar shapes for the couplings $g_j$, where there are the three distinct regions: (1) initial ramp, (2) flat bulk couplings and (3) emission ramp, as we show in Fig.~\ref{fig:comparison_true_exp}(a).
Furthermore, also the maximal clock precision deviates only negligably between the two cases as visualized in Fig.~\ref{fig:comparison_true_exp}(b).
This may not be too surprising since the optimization is initialized for small system sizes $n$ where we trust eq.\eqref{eq:gn_SM} to be the global optimum and then iteratively the ring length is increased $n\rightarrow n+1$ as well as the optimum of the previous $n$ is seeded as initial state for the optimization for $n+1$.
Since we expect the optimization landscape to change almost continuously from $n \rightarrow n+1$, we expect that our optimization remains close to the global optimum.

The optimal parameters for $\mu_\ell,\mu_r,g,\lambda_\ell$ and $\lambda_r$ are shown in the Fig.~\ref{fig:params_exp_scaling}.
We find that $\lambda_\ell \sim n^{0.35}$ scales approximately as our predicted power law $n^{1/3}$ from the main text.
Furthermore we find that $\lambda_r$ as well as $\mu_r$ do not scale with the ring size for $n\gg 1$.

One way to numerically solve the Lyapunov equation of this problem is to start by diagonalizing the effective Hamiltonian.
By writing $V^{-1}\left(-iH_\mathrm{eff}\right)V = \Lambda,$ where $\Lambda$ is a diagonal matrix, we can rewrite the Lyapunov recursion relation from eq.~\eqref{eq:rho_i_lyapunov} in the following way,
\begin{align}
    V^{-1}\rho_kV^\dagger = \Lambda V^{-1}\rho_{k+1} V^\dagger + V^{-1}\rho_{k+1}V^\dagger \Lambda^\dagger.
\end{align}
By defining $\sigma_k = V^{-1}\rho_k V^\dagger,$ the Lyapunov equation further reduces to
\begin{align}
\label{eq:sigma_i_recursive}
    \sigma_k = \Lambda \sigma_{k+1} + \sigma_{k+1}\Lambda^\dagger, 
\end{align}
which is solved in $O(n^2)$ steps because $\Lambda$ is diagonal.
It is possible to rewrite this equation element by element, $(\sigma_k)_{mn}=(\lambda_m + \lambda_n^*)(\sigma_{k+1})_{mn},$ which directly yields $\sigma_{k+1}$.
Given the diagonalization of $-iH_\mathrm{eff}$ with the right and left eigenvectors encoded in $V$ and $V^{-1}$ respectively, we can then recover $\rho_k = V\sigma_k V^{-1^{\dagger}}.$
In case we are interested in obtaining several moments of the waiting time distribution, we need to solve multiple iterations of the Lyapunov equation. 
This method has a better scaling because the standard Bartels-Steward Lyapunov solution method using a QR decomposition uses $O(n^3)$ steps~\cite{Bartels1972}. 
Particularly if $-iH_\mathrm{eff}$ can be efficiently diagonalized, using the recursive relation in~\eqref{eq:sigma_i_recursive} can be more efficient in terms of the computational runtime required compared to using a standard Bartels-Steward Lyapunov solver~\cite{Bartels1972}.

\section{Theory of clock precision in the limit of infinite entropy production} \label{SE:Ninf}
In this section we assemble all the arguments to understand one of the main results of this works: the scaling of precision at infinite entropy production with system size,
\begin{equation} \label{eq:NT_SM}
    \mathcal{N}_\infty \sim n^{4/3}
\end{equation}
as given in eq.~\eqref{eq:Nt_num_vs_thy} of the main text.
The waiting time distribution leading to this result sensitively depends on the propagation of a single localized excitation through the spin chain with couplings given in eq.~\eqref{eq:gn} of the main text.
Accordingly, the propagation can be divided into three parts; (1) preparation (see Sec.~\ref{SM:continuum_description}), (2) bulk propagation (see Sec.~\ref{SM:bulk_propagation_dispersion}) and (3) emission (see Sec.~\ref{SM:emission}),
which are discussed separately in the main text.
Combining all this leads to the desired scaling relation in eq.~\eqref{eq:NT_SM} and eq.~\eqref{eq:Nt_num_vs_thy} respectively, to which we supply the details in Sec.~\ref{SM:tick_moments}.

\subsection{\label{SM:continuum_description}Preparation}
The starting point of the system is the state $\ket{0}$, because after every tick, the system is reset to that state $\ket{0}$.
Without loss of generality we can chose the initial time as $t=0$.
This state then evolves according to $\ket{\psi(t)} = e^{-iH_0t}\ket{0}$ with the Hamiltonian
\begin{align}
\label{eq:H0_free}
    H_0 = \sum_{j=0}^{n-2} g_{j}\ketbra{j}{j+1} + \mathrm{h.c.}
\end{align}
Since we are only interested in the propagation through the initial ramp, we set $g_j = -\mu_\ell e^{-j/\lambda_\ell}+g$, neglecting the effects of the second ramp from
eq.~\eqref{eq:gn} and from the dissipator $J$, because $J^\dagger J \propto \ketbra{n-1}{n-1}$ acts only on the last site.
Contributions from $\overline J$ also play no role because we work in the limit of infinite entropy production for now.
Propagating the excitation through this exponential ramp will then prepare the wave packet subsequently travelling through the bulk of the ring which is discussed below in Sec.~\ref{SM:bulk_propagation_dispersion}.

Since the Hamiltonian in eq.~\eqref{eq:H0_free} possesses neither translational invariance nor any other structure (e.g. of T{\"o}plitz form) allowing for a closed solution we will instead tread it approximately in the continuum with the hydrodynamic limit.
To that end, large rings and soft ramps have to be considered $n, \lambda_\ell \gg 1$, where the wave functions barely varies on the scales of a single lattice spacing.
This allows us to find an approximate continuum description of the Schr{\"o}dinger equation on the lattice.
In this limit all effects which are attributed to the lattice are being neglected.
Writing the Schrödinger equation $i\partial_t \ket{\psi(t)}=H_0\ket{\psi(t)}$ element by element we obtain
\begin{align}
\label{eq:partial_t_psi_n}
    i\partial_t \psi_j(t) = g_j\psi_{j+1}(t) + g_{j-1}\psi_{j-1}(t),
\end{align}
where $\psi_j(t)=\braket{j}{\psi(t)}$ is the amplitude of the wave function at site $j$.
This equation is simplified by the canonical transformation $|\tilde\psi(t)\rangle=U\ket{\psi(t)}$, where 
\begin{align}
\label{eq:U_trsf}
    U = \sum_{j=0}^{n-1} e^{i\frac{j\pi}{2}}\ketbra{j}{j}.
\end{align}
Elementwise, the transformation reads $\tilde\psi_j(t)=i^j \psi_j(t)$ which is reminiscent of targeting the hydrodynamic mode in relativistic Dirac fermions in one dimension which yields the Luttinger liquid \cite{Fradkin_13,Shankar_17}. 
This canonical transformation renders the Schrödinger equation~\eqref{eq:partial_t_psi_n} completely real
\begin{align}
\label{eq:partial_phi_n}
    \partial_t \tilde\psi_j(t) &= -g_j\tilde\psi_{j+1}(t) + g_{j-1}\tilde\psi_{j-1}(t).
\end{align}
Note that the fully localized initial condition $\psi_j(0)=\delta_{j,0}$ is left untouched by this transformation $\tilde\psi_j(0)=\delta_{j,0}$.
Since the initial condition is real so is $\tilde\psi_j(t)$ for all times $t\ge 0$.
For simplicity, we will set $\tilde{\psi}_j \rightarrow \psi_j$ in the following.
In order to perform the continuum limit we consider an equidistant lattice at $x_j = {j}/{\lambda_\ell}$ on the real line, where $j=0,\dots,n-1$.

The continuum limit of the discrete lattice wave function is done by finding continuous functions $\Psi(t,x)$ and $g(x)$ such that
\begin{align}
\label{eq:discrete_continuous_identification}
    \psi_j(t) \equiv \frac{1}{\sqrt{\lambda_\ell}}\Psi(t,x_j),
    \text{ and }
    g_j\equiv g(x_j)
\end{align}
coincide.
In the limit $\lambda_\ell, n\rightarrow \infty$, this can be achieved by identifying the continuum generalization $g(x) = -\mu_\ell e^{-x} + g$ for the couplings, and hydrodynamic limit of the Schrödinger equation in eq.~\eqref{eq:partial_phi_n} giving
\begin{align}
\label{eq:partial_phi_continuum}
    \partial_t \Psi(t,x) = -\frac{1}{\lambda_\ell}\left(2g(x)\partial_x\Psi(t,x) + (\partial_x g(x))\Psi(t,x)\right).
\end{align}
To check the consistency of this expression with the discrete equation we first identify the finite differences in the discrete case with partial derivatives in the continuum case,
\begin{align} \label{eq:partial_ident_1}
    \partial_x g(x_j) = \frac{g_{j+1}-g_j}{\lambda_\ell} + O\left(\lambda_\ell^{-2}\right),
\end{align}
and,
\begin{align}  \label{eq:partial_ident_2}
    \partial_x \Psi(x_j,t)=\frac{\psi_{j+ 1}(t) -\psi_j(t)}{\lambda_\ell} + O(\lambda_\ell^{-2}).
\end{align}
Inserting the identifications of eqs.~\eqref{eq:partial_ident_1} and~\eqref{eq:partial_ident_2} into eq.~\eqref{eq:partial_phi_continuum} recovers eq.~\eqref{eq:partial_phi_n} up to terms of order $O(\lambda_\ell^{-2})$.
As for the normalization in the continuum $1=\int_0^\infty dx |\Psi(x,t)|^2$, going to a Riemann sum shows it is compatible with the normalization of the discrete wave function $\sum_{j=0}^{n-1} |\psi_j(t)|^2 = \sum_{j=0}^{n-1}\frac{1}{\lambda_\ell }|\Psi(t,x)|^2\rightarrow 1,$ in the limit of $n,\lambda_\ell \gg 1$.

So long as both $\Psi(t,x)$ and $g(x)$ vary slowly on the length scale of the lattice, this approximation is good and the error can be controlled with the limit $\lambda_\ell\gg 1$.
We can thus approximate the evolution of $\ket{\psi(t)} = \sum_j \psi_j(t)\vert j\rangle$ using the continuum limit $\Psi(t,x)$.
In order to assess whether $\Psi(t,x)$ may be really interpreted as a wave function even for finite $\lambda_\ell$ and $n$ we have to ensure that $n(t,x) = \vert \Psi(t,x)\vert^2$ is a probability density and that $\int_0^\infty \mathrm{d}x\,n(t,x) =1$ remains normalized exactly for all times.
Using eq.~\eqref{eq:partial_phi_continuum} we derive an effective equation of motion for $n(t,x)$ which gives
\begin{equation}
    \partial_t n(t,x) =  -\frac{1}{\lambda_\ell} \partial_x \left(2g(x)n(t,x)\right).\label{eq:partial_rho_continuum}
\end{equation}
This equation has the form of a transport equation~\cite{Evans22partial} well studied in hydrodynamics. 
Since the left hand side is a total derivative with respect to $x$, integrating both sides shows that the integral of $n(t,x)$ will not change and that $\Psi(t,x)$ can indeed be treated as a wave function and eq.~\eqref{eq:partial_phi_continuum} is a valid Schrödinger equation.

Both the equations for $\Psi(t,x)$ in~\eqref{eq:partial_phi_continuum} and for $n(t,x)$ in~\eqref{eq:partial_rho_continuum} can be solved analytically using the method of characteristics~\cite{Evans22partial}, a standard tool for studying first order partial differential equations (PDE).
Here, we focus on the evolution of the probability density $n(t,x)$, which is the relevant quantity describing the transport of the wave packet and subsequently the waiting time distribution.
For the analytical solution it is advantageous to expand the derivative  equation~\eqref{eq:partial_rho_continuum} which gives
\begin{align}
\label{eq:partial_rho_continuum_full}
    \partial_t n(t,x) + \frac{2}{\lambda_\ell} g(x) \partial_x n(t,x) = - \frac{2}{\lambda_\ell} (\partial_x g(x))n(t,x).
\end{align}
This is solved analytically by the characteristic
\begin{align}
    \xi(t,x) = \log\left(\frac{\mu_\ell}{g}
                    \left(1 - e^{-\frac{2gt}{\lambda_\ell}}\right)
                    + e^{x - \frac{2gt}{\lambda_\ell}}\right),
\end{align}
and expressing eq.~\eqref{eq:partial_rho_continuum_full} as a function of $(t,\xi)$ instead of $(t,x)$. 
We then relate the total $t$-derivative of the density $n$ expressed in $(t,\xi)$ coordinates to eq.~\eqref{eq:partial_rho_continuum_full} by making the identification $\partial_t x(t,\xi) = \frac{2}{\lambda_\ell} g(x(t,\xi))$.
The PDE for $n(t,x)$ then turns into an ordinary differential equation that can be solved exactly.
For an initial condition $n(0,x) = p(x)$ whose support is constrained to the positive reals $\mathbb{R}_{+}$, we get the following exact solution
\begin{align}
\label{eq:rho_x_t_exact}
    n(t,x) = \frac{p(\xi(t,x))}{1 + \frac{\mu_\ell}{g}\left(e^{\frac{2gt}{\lambda_\ell}-x} - e^{-x}\right)}.
\end{align}
With this solution we can study the asymptotic state $n(t,x)$ long after it has left the inital ramp $2gt \gg \lambda_\ell$. 
In this limit, we obtain
\begin{align}
    \xi(t,x) &= \log\frac{\mu_\ell}{g} + \log\left(1 + \frac{g}{\mu_\ell}e^{x - \frac{2gt}{\lambda_\ell}}\right) + O\left(e^{-\frac{2gt}{\lambda_\ell}}\right).
\end{align}
and the denominator in~\eqref{eq:rho_x_t_exact} becomes a function of $x-{2gt}/{\lambda_\ell}$.

We continue by considering the real-space asymptotics of the solution in eq.~\eqref{eq:rho_x_t_exact} in the regime $x\gg 2gt/\lambda_\ell$ and $x\ll 2gt/\lambda_\ell$, still for times where the wave packet has left the initial ramp.
We find that for large values of $x$, the initial distribution $p(x)$ of the wave packet dominates, because $\xi(t,x) = x-2gt/\lambda_\ell + O(e^{-(x-2gt/\lambda_\ell)})$, and the denominator goes to $1$ in that limit.
For small values of $x$, on the other hand, we find $\xi(t,x) = \log \frac{\mu_\ell}{g} + O(e^{-(2gt/\lambda_\ell-x)})$ goes to a constant and the exponential from the denominator dominates.
Both together yield the asymptotics of
\begin{align}
n(t,x) = \left\{
\begin{matrix}
&p\left(x-\frac{2gt}{\lambda_\ell}\right),\qquad &x\gg \frac{2gt}{\lambda_\ell},\\
& \frac{g}{\mu_\ell}p\left(\log \frac{\mu_\ell}{g}\right)e^{-\left(\frac{2gt}{\lambda_\ell}-x\right)}, \qquad & x\ll \frac{2gt}{\lambda_\ell}.
\end{matrix}\right.
\end{align}
Therefore, asymptotically, the wave packet decays exponentially on its back side pointing away from the direction of propagation.
The asymptotic behavior on its front side towards the direction of propagation is given by the shape of the initial state $p(x)$.

Finally we study the eq.~\eqref{eq:rho_x_t_exact} in the limit where $2gt$ is comparable with $\lambda_\ell$ as well as $x \gg 1$.
In this limit the solution can be cast in the form
\begin{align}
\label{eq:rho_x_t_scaling}
    n(t,x) = f\left(x-\frac{2gt}{\lambda_\ell}\right)^2 + O\left(e^{-\frac{2gt}{\lambda_\ell}}\partial_x p\right),
\end{align}
with some function $f$ dependent on the initial state density $p$ which corresponds to the wave function $\Psi(t,x) = f(x-2gt/\lambda_\ell)$.
The initial state on the lattice is perfectly localized in $\vert 0 \rangle$ in real space.
An infinitely localized state as an initial condition is of course at odds with our effective hydrodynamic description but we will initialize a wave packet in a $p(x)$ which is strongly concentrated on hydrodynamics scales (for example a Gaussian).
While this treatment is very crude, the numerics in Fig.~\ref{fig:width_scaling}(a) suggest that the relevant quantity below, the width, is well captured in this approach, even though the hydrodynamic approximation looses quickly oscillating effects on the lattice length scale.
This is the main result of this section is the preparation of the wave packet in the form~\eqref{eq:rho_x_t_scaling}.
This form is later crucial to obtain the scaling of the clock precision.

\subsection{\label{SM:bulk_propagation_dispersion}Bulk propagation}
In Sec.~\ref{SM:continuum_description}, we studied the asymptotic form of the wave packet long after it has left the initial ramp using a hydrodynamics description.
We will now analyze how the wave packet propagates through bulk of the ring. 
This limit we describe in terms of the effective translationally invariant Hamiltonian 
\begin{align}
\label{eq:H_bulk}
    H_\mathrm{bulk} = \sum_{j=0}^{n-1} g\ketbra{j+1}{j}+\mathrm{h.c.},
\end{align}
in good agreement with $g_j$ from eq.~\eqref{eq:gn} in the limit where $n -\lambda_r \gg j \gg\lambda_\ell$. 
Leveraging the translational invariance of eq.~\eqref{eq:H_bulk}, we go from a basis of localized states on the lattice $\ket{j}$ to momentum space $\ket{\psi_{k_\ell}}$. 
This is given by
\begin{align}
    \label{eq:ket_theta_k}
    \ket{\psi_{k_\ell}}:=\frac{1}{\sqrt
    n}\sum_{j=0}^{n-1}e^{-ik_\ell j} \ket{j},
\end{align}
where $k_\ell = {2\pi \ell}/{n}$ is the discrete lattice momentum for $\ell=0,\dots,n-1$.
The momentum eigenstates $\ket{\psi_{k_\ell}}$ from~\eqref{eq:ket_theta_k} diagonalize $H_\mathrm{bulk}$ with dispersion relation (eigenvalues) given by $E(k_\ell) = 2g\cos(k_\ell)$.
The coefficient of the wave packet $\ket{\psi(t)}$ in momentum space $\psi_{k_\ell}(t) = \braket{\psi_{k_\ell}}{\psi(t)}$ are
\begin{align}
    \psi_{k_\ell}(t) &= \frac{1}{\sqrt{n}}\sum_{j=0}^{n-1}e^{ik_\ell j}\psi_j(t).
\end{align}

While the continuum solution in Sec.~\ref{SM:continuum_description} was used to capture the rapid preparation process of the wave packet, it falls short to capture dispersive effects which occur when traveling for longer times.
To get a better grasp on this, let us evolve the wave function up to some time $t_0$ which was chosen such that the wave packet has just left the initial ramp, $2gt_0\gtrsim \lambda_\ell$.
For times $t\geq t_0,$ the evolution is well described in terms of the effective bulk Hamiltonian~\eqref{eq:H_bulk}.
Numerically (see Fig.~\ref{fig:RK_space}), we see that $\psi_{k_\ell}(t_0)$ is super-exponentially concentrated around $k_0=\pi/2$ in momentum space, and therefore we can write the evolution for times later than $t_0$ as
\begin{align}
\label{eq:psi_n_t0+t}
    \psi_j(t_0+t) &= \frac{1}{\sqrt{n}}\sum_{\ell=0}^{n-1}\psi_{k_0 +q_\ell}(t_0) e^{-2igtE(k_0+q_\ell)}e^{-i(k_0+q_\ell)j},
\end{align}
where we set $k_\ell= k_0 + q_\ell$.
The phase profile $e^{i\pi j /2} = i^j$ is the same we encountered in eq.~\eqref{eq:U_trsf} and corresponds to the rapidly oscillating phase profile on top of a slowly varying wave function.
Since we will later only be interested in absolute squares of the wave function this will not contribute and we will drop it below.
Due to the strong concentration of $\psi_{k_\ell}(t_0)$ only the parts of the dispersion relation close to $k_0$ will contribute to the dynamics and we expand
\begin{align}
    E(k_0+q) = -2gq +\frac{2gq^3}{3!} + O(gq^5),
\end{align}
allowing us to rewrite~\eqref{eq:psi_n_t0+t} as
\begin{align}
\label{eq:psi_n_t0+t_expanded}
    \psi_j(t_0+t) = \frac{1}{\sqrt{n}}\sum_{\ell=0}^{n-1}\psi_{k_0+q_\ell}(t_0) e^{-i q_\ell (j - 2gt)}e^{-\frac{i}{3}gt q_\ell^3},
\end{align}
where we did not write the terms of order $O(q_\ell^5)$ and higher due to the strong concentration of $\psi_{k_\ell}(t_0)$ around $k_0$.
The term with first order is responsible for translating the wave packet to the right with group velocity $2g$.
Broadening of the wave packet is caused by the third-order term.
Its consequences are the main concern of this section.

To assess how exactly the wave function, prepared by the initial ramp, traverses the bulk part of the ring clock we need to know the shape of the wave function at $t_0$.
An exact treatment may be cumbersome analytically and give little additional insight.
For our aim of determining the clock's ticking statistics, the scaling behavior from the hydrodynamic solution obtained in the continuum, turns out to be sufficient,
\begin{align}
\label{eq:psi_x_n_f_n}
    \psi_j(t) = \frac{1}{\sqrt{\lambda_\ell}}\Psi(t,x_j) \sim \frac{1}{\sqrt{\lambda_\ell}}f\left(\frac{j-2gt}{\lambda_\ell}\right),
\end{align}
where we have set $x_j =  j/\lambda_\ell$ in eq.~\eqref{eq:rho_x_t_scaling}.

We will now consider the wave function of eq.~\eqref{eq:psi_x_n_f_n} in momentum space with respect to the discrete lattice momenta which gives us
\begin{align} 
    \psi_{k_0+q_\ell}(t_0) &= \frac{1}{\sqrt{n}} \sum_{j=0}^{n-1} e^{i(k_0+q_\ell) j} \psi_j(t_0) \\
    &\sim \sqrt{\frac{\lambda_\ell}{n}}\int \mathrm{d}x e^{i q_\ell \lambda_\ell x} f\left(x - \frac{2gt_0}{\lambda_\ell}\right) \\
    &= \sqrt{\frac{\lambda_\ell}{n}}\hat{f}(q_\ell \lambda_\ell),
\end{align}
up to an irrelevant phase-shift due to the initial translation $2gt_0/\lambda_\ell$ in the final line.
We replaced the Riemann sum $\sum_j = \lambda_\ell \sum_x \Delta x \rightarrow \int\mathrm{d}x$ with $\Delta x = 1/\lambda_\ell$ by an integral in the second line in the limit $\lambda_\ell\gg 1$. 
While we are not privy to the exact form of $f$, it is still possible to formally obtain its Fourier transformation $\hat{f}$, which is also strongly localized.
From the dependence of the $\hat{f}$ on the length of the ramp $\lambda_\ell$ we conclude that the longer the initial ramp the stronger the localization in momentum space as found in Fig.~\eqref{fig:RK_space}.

Now that we have some notion of a state after the initial ramp, we are in the position to study the propagation of the wave packet using eq.~\eqref{eq:psi_n_t0+t_expanded}.
In the limit of large $n \gg 1$, the discrete sum over $k_\ell\in \frac{2\pi}{n}\mathbb Z_n$ becomes an integral over $k\in [0,2\pi)$ from which we obtain
\begin{align}
    \psi_j(t_0+t) &= \frac{\sqrt{\lambda_\ell}}{n}\sum_{\ell=0}^{n-1}\hat{f}(q_\ell \lambda_\ell) e^{-iq_\ell \lambda_\ell \frac{j-2gt}{\lambda_\ell}} e^{-i\frac{2gt}{\lambda_\ell^3} \frac{(q_\ell \lambda_\ell)^3}{3!} } \nonumber \\
    &\sim \sqrt{\lambda_\ell} \int_{-\infty}^\infty \frac{\mathrm{d}Q}{2\pi} \hat{f}(Q) e^{-iQ \frac{j-2gt}{\lambda_\ell}} e^{-i\frac{2gt}{\lambda_\ell^3} \frac{Q^3}{3!}},\label{eq:correction_q3_explicit}
\end{align}
Note that the length of an $k$-space interval is $\Delta k = 2\pi/n$ so the sum can be written as $\frac{1}{\Delta k}\sum_j \Delta k$ as a Riemann sum.
In the limit $n\gg 1$ we replace the sum by an integral and obtain $\frac{2\pi}{n}\sum_{j=0}^{n-1}\rightarrow\int_{-\pi/2}^{3\pi/2} \mathrm{d}q$, where we have again used the translational invariance of the dispersion relation.
Finally, we rescale the momentum $Q=q\lambda_\ell$ and since the $\hat{f}$ is strongly localized im momentum space, extending the limits of the integral to infinity will not incur too much of an error. 

We now analyze the broadening of the wave packet due to the cubic term in the exponent in eq.~\eqref{eq:correction_q3_explicit}.
Up to numerical constants the prefactor scales like $2gt/\lambda_\ell^3$ and therefore the longer the wave propagates the more relevant become the dispersive effects.
This very observation is the main result of this section and crucial to finding the length of the ramp $\lambda_\ell$ which guarantees propagation with the least amount of spreading. 

\subsection{\label{SM:emission}Emission}
In the following two subsections, we first discuss in Sec.~\ref{SM:apodization} the physics of the tick emission, investigating how the choice of couplings on the right ramp of Fig.~\ref{fig:visualization}(d) in the main text allows for the excitation to be emitted with unit probability once it arrives at the final site.
Then, in the following Sec.~\ref{SM:tick_moments}, we analyze the probability distribution of when the excitation is emitted based on the coherent evolution in the bulk of the ring clock, which allows us to obtain the scaling relations $\mathrm{E}[T]\sim n$ and $\mathrm{Var}[T]\sim n^{2/3}.$

\subsubsection{\label{SM:apodization}Apodization of the right ramp}
\begin{figure*}
    \centering
    \includegraphics[width=\textwidth]{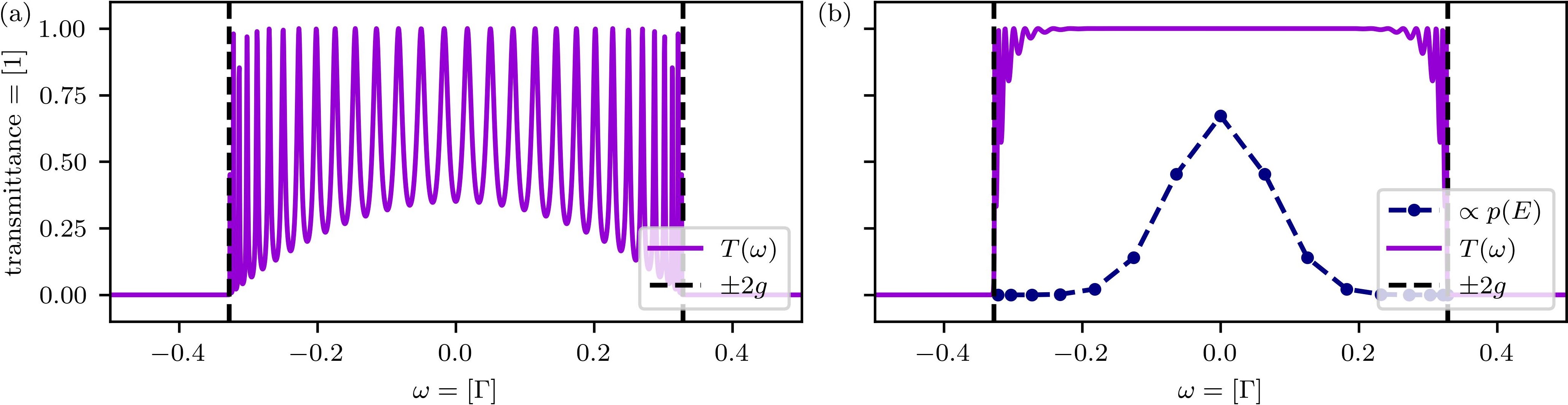}
    \caption{Here, we show the transmission function of our setup. In (a) with flat couplings, and in (b) with apodized couplings, both for $n=32$ sites.
    The transmission function $T(\omega)$ can be calculated using the expression in eq.~\eqref{eq:T(E)} by understanding the ring setup as a symmetric scatterer. 
    In (a) we show the transmission function in the case where all nearest neighbor couplings $g_j=g$ are constant.
    The Fabry-Perot interference fringes prevent a unit probability emission of any spatially located travelling wave packet, due to the non-unity transmission function~\cite{Chak2006}.
    (b) Here, we have the transmission probability for the case where the couplings at the both ends of the scatterer are symmetrically apodized with a ramp as in the ring clock. Furthermore, we plot the energy probability distribution of the wave packet in the bulk of the ring, rescaled for better readability.
    We see that this setup has an approximately unit transmittance in the energy range where the wave packet has support.
    In both panels, the vertical dashed lines show the ends of the transmission spectrum given by the dispersion relation $E=2g\cos(k) \in [-2g,2g]$.}
    \label{fig:transmittance}
\end{figure*}
The choice of the couplings $g_j$ at the end of the ring cycle can be understood as a boundary matching problem, separate from the initialization of the wave packet at the beginning of the evolution.
By maximizing for the clock precision $\mathcal N_\infty,$ we find that the last few couplings $g_{n-2},g_{n-3},g_{n-4},\dots$ are larger than the coupling $g$ in the bulk of the ring, as visualized in Fig.~\ref{fig:visualization}(d) of the main text.
The increase in the couplings is reminiscent of the optimal values used in the apodization of resonant tunneling structures to maximize transmission in a transport setting~\cite{Sumetsky2003,Chak2006,ferreira2021}.
To have a unit transmission coefficient, it has been shown that only a small constant number of couplings before the emitting site have to be adjusted, irrespective of how large the bulk is.
Such an apodization of the final couplings prevents the reflection of most travelling waves within a certain energy band, and as it turns out,
our maximization of clock precision leads to a choice of couplings that favor transmission over reflection (see comparison Fig.~\ref{fig:transmittance}(a,b)).

\paragraph*{Transmission probability.} To make a more quantitative connection between tick emission in the ring clock and transimission in a quantum transport setting, one may wish to think of an open chain instead of a ring.
Since for the problem of emission only the final sites are relevant, the initialization ramp discussed in Sec.~\ref{SM:continuum_description} can be ignored.
For quantifying the transmission behavior of the right ramp, we look at a symmetric setting, where the couplings on the left are the same as those on the right, that is, $g_j^{\rm sym.} = g_{n-2-j}^{\rm sym.}.$
With our model as given in eq.~\eqref{eq:gn}, the couplings of the symmetrized setting are then described by $g_j^{\rm sym.}=g + \mu_r\left(e^{-j/\lambda_r} + e^{-(n-j-1)/\lambda_r}\right)$.
This setup allows for probing the emission ramp on the right independent of the details of wave packet preparation.
Combining the resulting transmission function with the results on the travelling wave packet from our previous analysis in Sec.~\ref{SM:bulk_propagation_dispersion}, we find that the excitation lies precisely within the energy spectrum that our choice of couplings transmits, as shown in Fig.~\ref{fig:transmittance}(b).
Transmittivity can be quantified using the non-equilibrium Green's function (NEGF) method, for example, and in the following derivation we follow Ref.~\cite{Datta2005a}.
The transmission probability for a travelling wave of energy $\omega$ in our setting is given by
\begin{align}
\label{eq:T(E)}
    T(\omega) = \Gamma^2\left|\left\langle n-1 \left|(\omega - H_\mathrm{eff})^{-1}\right| 0 \right\rangle\right|^2,
\end{align}
where the effective Hamiltonian is defined as in Sec.~\ref{SM:calculating_precision}, but with the symmetrized couplings, i.e.,
\begin{align}
\label{eq:H_eff_symm}
    H_\mathrm{eff} &= \sum_{j=0}^{n-2} g_j^{\rm sym.}\ketbra{j+1}{j} + \mathrm{h.c.} \nonumber\\
    &\quad - i\frac{\Gamma}{2}\ketbra{n-1}{n-1} - i\frac{\Gamma}{2}\ketbra{0}{0}.
\end{align}
We show a comparison between how the transmission function looks like for different choices of couplings.
In Fig.~\ref{fig:transmittance}(a) we show the case without apodization by using flat couplings on the entire chain.
In Fig.~\ref{fig:transmittance}(b) we show the transmission function for the case where the ends of the chain are apodized using the optimal couplings from the ring clock.
Even though we optimize for clock precision and not for transmittivity directly, this analysis shows that the optimal choice of couplings produces a travelling wave packet in the bulk of the ring whose energy distribution lies exactly within the transmission window of the apodized region at the end of the ring. 

\paragraph*{The non-equilibrium Green's function.}
The NEGF method is an exact, non-perturbative method to determine scattering amplitudes and for the following derivation we adapt the techniques from~\cite{Datta2005a} to our context.
In our setting it is sufficient to look at a system comprising three parts, a left port, the central region (the chain), and the right port.
The left and right ports are microscopic models for the baths from which excitations enter and leave the chain.
For the NEGF method, we restrict to the one-particle subspace $\mathcal H_L^1 \oplus \mathcal H_C^1 \oplus \mathcal H_R^1 \subseteq \mathcal H_L\otimes\mathcal H_C\otimes\mathcal H_R$ in the full Hilbert-space, as in the main text.
As a consequence, we can write the Hamiltonian describing the dynamics using
\begin{align}
\label{eq:H_total}
    H = 
\begin{bmatrix}
H_{LL} & H_{LC} & 0 \\ 
H_{CL} & H_{CC} & H_{CR}\\ 
0 & H_{RC} & H_{RR}
\end{bmatrix}.
\end{align}
The Green's function must, by definition, satisfy the condition $(\omega \pm i0 - H)G(\omega)^\pm  = \mathds 1,$ where the notation $\pm i0$ can be understood as $\pm i\varepsilon$ in the limit of $\varepsilon\rightarrow 0^+$ in a distribution sense, to regularize the Green's function.
In the following, we only work with the retarded Green's function, that is, the limit $+i0$ to ensure causality (see e.g. Chapter~8 from~\cite{Datta2005a}), and we will drop the additional superscript and simply write $G(\omega)$ to mean $G(\omega)^+$.
It will be useful to consider the following examples: the Green's function of the isolated leads are given by $g_L(\omega) = \left(\omega + i0 - H\right)^{-1}$ for the left lead and similarly for the right lead. Furthermore, the central element of the total Green's function for $H$ as defined in~\eqref{eq:H_total} is the inverse of the effective Hamiltonian, $\left(\omega - H_\mathrm{eff}\right) G_{CC}(\omega) = \mathds 1,$ where the effective Hamiltonian in this general setting is given by
\begin{align}
\label{eq:H_eff_NEGF}
    H_\mathrm{eff}(\omega) = H_{CC} + H_{CR} g_R(\omega) H_{RC} + H_{CL} g_L(\omega) H_{LC}.
\end{align}
This can be derived by solving the three equations in the middle column of $(\omega+i0-H)G(\omega)=\mathds 1$ for $G_{CC}(\omega).$
The self-energies $\Sigma_{L}(\omega) = H_{CL}g_L(\omega) H_{LC}$ (similarly for $R$) arise from the coupling of the central region to the leads and they can be split into a real part (the Lamb-shift) and an imaginary part, which is the non-Hermitian term describing the loss or gain of excitations due to the coupling of the central region to the ports.
Generally, the imaginary term is identified as the sink term modeling particle loss (or gain) due to the coupling with the port, $\mathfrak{Im}[\Sigma_L(\omega)] \equiv \Gamma_L(\omega)/2$.
By using the Shokhotski-Plemelj formula~\cite{Sokhotskii1873,Plemelj1964} stating $(x\pm i0)^{-1} = \mathcal P x^{-1} \mp i\pi \delta(x)$ with $\mathcal P$ denoting the Cauchy-principle value, we can write
\begin{align}
    \Gamma_L(\omega) &= -2\pi H_{CL}\delta(\omega - H_{LL})H_{LC} \\
    &= -iH_{CL}(g_L(\omega) - g_L(\omega)^\dagger) H_{LC}
\end{align}
While the details of this expression are irrelevant to our purpose, one could in principle use them to exactly solve the scattering problem at hand.
We, however, work in an approximate limit, where the imaginary term of the self-energies of the left and right port simplify to the expressions we have been using so far, where for the energy range $\omega$ of interest, we assume that
\begin{align}
\label{eq:Gamma_port_def_L}
    \mathfrak{Im}[\Sigma_{L}(\omega)] \equiv \frac{\Gamma_L(\omega)}{2} = -\frac{\Gamma}{2}\ketbra{0}{0}_C, 
\end{align}
and,
\begin{align}
\label{eq:Gamma_port_def_R}
    \mathfrak{Im}[\Sigma_{R}(\omega)] \equiv \frac{\Gamma_R(\omega)}{2} = -\frac{\Gamma}{2}\ketbra{n-1}{n-1}_C.
\end{align}
Such a model can be recovered using a Breit-Wigner -- or Markov -- approximation~\cite{Landau1977}, as already implicitly done in the main text by using the Lindblad master equation.
By furthermore absorbing the real part of the self-energy in the Hamiltonian $H_{CC}$ and shifting the diagonal of the chain Hamiltonian to $0,$ we recover the expression in eq.~\eqref{eq:H_eff_symm} that we have already been using in the main part of the paper (up to the symmetrized setting).

\paragraph*{Derivation of the transmission probability.}
Given the consistency with the previous model, i.e., equality of $H_\mathrm{eff}$ as in~\eqref{eq:H_eff_symm} and as derived using the NEGF method~\eqref{eq:H_eff_NEGF}, we are ready to move further towards deriving the transmission probability $T(E)$ from eq.~\eqref{eq:T(E)}.
To this end, we look at the eigenvalue equation of the full Hamiltonian including both ports and the chain region,
\begin{align}
\label{eq:(omega-H)Psi}
    \left(\omega - H\right) \ket{\Psi} = 0,\text{ where } \ket{\Psi} :=
    \begin{bmatrix}
\ket{\phi_L} + \ket{\chi_L}\\ 
\ket{\psi_C}\\ 
\ket{\chi_R}
\end{bmatrix}.
\end{align}
Physically, we split the wave function on the left port into one incoming contribution $\ket{\phi_L}$ and a back-scattered contribution $\ket{\chi_L}.$
Furthermore, we have the wave function $\ket{\psi_C}$ in the chain and the transmitted function $\ket{\chi_R}$ in the right port.
Given the incoming wave function $\ket{\phi_L}$ as the independent variable, we want to determine the responses $\ket{\chi_L},\ket{\chi_R}$ and $\ket{\psi_C}.$
Note that none of these terms is necessarily normalized, we only chose $\ket{\phi_L}$ to be normalized by convention.
Furthermore, we assume $(\omega - H_{LL})\ket{\phi_L}=0,$ that is, we chose the incoming wave to have energy $\omega,$ then, we can solve the time-independent Schrödinger equation~\eqref{eq:(omega-H)Psi} for $\ket{\chi_L}$ and $\ket{\chi_R}$.
We present the examplary case for $\chi_L$, by looking at the first row,
\begin{align}
    (\omega - H_{LL})(\ket{\phi_L} + \ket{\chi_L}) - H_{LC}\ket{\psi_C} = 0,
\end{align}
from which it follows by using $(\omega-H_{LL})\ket{\phi_L}=0$ that $(\omega-H_{LL})\ket{\chi_L} = H_{LC}\ket{\psi_C}$.
Formally, we can only invert this equation by adding a regularizing imaginary part to the energy $\omega \mapsto \omega +i0,$ which then allows us to invert this equation to give $\ket{\chi_L} = g_L(\omega) H_{LC}\ket{\psi_C}$.
In the literature, as e.g.\ Ref.~\cite{Datta2005a}, this problem is usually solved by adding an additional source-term $\ket{S}$ to the time-independent Schrödinger equation $(\omega +i\varepsilon -H)\ket{\Psi} = \ket{S}$, which then vanishes in the limit $\varepsilon\rightarrow 0$.
For the right response we can similarly write $\ket{\chi_R} = g_R(\omega)H_{RC}\ket{\psi_C}$ using the last row of eq.~\eqref{eq:(omega-H)Psi}.
Inserting this into the middle row gives us an expression for $\ket{\psi_C}$ as a function of $\ket{\phi_L},$
\begin{align}
\label{eq:(omega_Heff)psi=H_CL_phi}
    (\omega - H_\mathrm{eff})\ket{\psi_C} = H_{CL}\ket{\phi_L}.
\end{align}
Using the same regularization from before $\omega\rightarrow\omega+i0$ allows us to formally solve for the wave function in the chain $\ket{\psi_C} = G_{CC}(\omega)H_{CL}\ket{\phi_L}$,
giving us the expression for the scattered wave-function in the right port as a function of the incoming wave $\ket{\chi_R} = g_R(\omega)H_{RC} G_{CC}(\omega) H_{CL}\ket{\phi_L}$.
Generally, there are many possible left-port solutions $\ket{\phi_L^k}$ for the incoming wave, and for later use, we will denote these states with the superscript $k$. 
Given the relationship between $\ket{\chi_R}$ and $\ket{\phi_L^k}$, we can determine the probability current from the left port to the right port using the time-dependent Schrödinger equation $i\partial_t \ket{\Psi} = H \ket{\Psi},$ noting that $\ket{\Psi}$ is still the solution of the stationary equation~\eqref{eq:(omega-H)Psi}.
We are interested in the probability current into the right port, for which we need to calculate $i_T^k = \partial_t \braket{\chi_R}{\chi_R}$, with the subscript $T$ for the transmitted current, and the superscript $k$ for indexing the implicit incoming state $\ket{\phi_L^k}$.
One may think that because $\ket{\Psi}$ is the solution to the stationary Schrödinger equation, the probability current is zero; here, this is not the case, because we look at the perturbed equation where $\omega\mapsto\omega +i0$, thus giving us the non-zero current,
\begin{align}
    i_T^k &= -i\bra{\chi_R}\left(H_{RC}\ket{\psi_C} + H_{RR}\ket{\chi_R}\right) + \mathrm{h.c.} \\
    &= -i\left(\braket{\chi_R}{H_{RC}|\psi_C} - \braket{\psi_C}{H_{CR}|\chi_R}\right)
\end{align}
Using the previous result that $\ket{\chi_R}=g_R(\omega)H_{RC}\ket{\psi_C}$ together with the result that the imaginary part of the self energy equals the non-Hermitian loss term $\mathfrak{Im}[\Sigma_R(\omega)]=\Gamma_R(\omega)$, we can simplify
\begin{align}
    i_T^k &= i\braket{\psi_C}{H_{CR}\left(g_R(\omega) - g_R(\omega)^\dagger\right)H_{RC}|\psi_C} \\
    &= -\braket{\psi_C}{\Gamma_R|\psi_C},
\end{align}
where we dropped the $\omega$ argument in the last line for readability. We will from now on not explicitly write the $\omega$ argument anymore in the cases where it is unambiguous.
By using the relationship between $\ket{\psi_C}$ and the incoming wave $\ket{\phi_L^k}$ derived in eq.~\eqref{eq:(omega_Heff)psi=H_CL_phi}, we can write the current as a function of the incoming wave $\ket{\phi_L}$.
The transmitted current expressed as a function of the incoming wave $\ket{\phi_L^k}$ is thus given by,
\begin{align}
    i_T^k &= -\braket{\phi_L^k}{H_{LC}G_{CC}^\dagger \Gamma_R G_{CC}H_{CL}|\phi_L^k} \\
    &= -\tr\left[H_{CL}\ketbra{\phi_L^k}{\phi_L^k} H_{LC}G_{CC}^\dagger \Gamma_R G_{CC}\right].
\end{align}
Summing over all possible eigenstates $\ket{\phi_L^k}$ with energy $\omega_{L,k}=\omega$ and occuppation probability $p(\omega_k)$ we obtain the total transmitted probability current,
\begin{align}
    i_T &= \sum_{k} p(\omega_k) i_T^k  \\
    &= \int d\omega p(\omega) \sum_k \delta(\omega-\omega_{L,k}) i_T^k \\
    &= \int \frac{d\omega}{2\pi}p(\omega)\tr\left[\Gamma_L G_{CC}^\dagger \Gamma_R G_{CC}\right],
\end{align}
from which we can extract the transmission probability
\begin{align}
    T(\omega)=\tr\left[\Gamma_L G_{CC}^\dagger \Gamma_R G_{CC}\right]
\end{align}
as in~\cite{Datta2005a}.
Inserting the operators from our model~\eqref{eq:Gamma_port_def_L} and~\eqref{eq:Gamma_port_def_R}, $\Gamma_L=\Gamma \ketbra{0}{0}$ and $\Gamma_R=\Gamma\ketbra{n-1}{n-1}$, we recover the expression $T(\omega)=\Gamma^2|\braket{0}{G_{CC}(\omega)|n-1}|^2$ from eq.~\eqref{eq:T(E)} at the beginning of this section.

\subsubsection{\label{SM:tick_moments} Scaling of clock precision with ring length}
\begin{figure*}
    \centering
    \includegraphics[width=\textwidth]{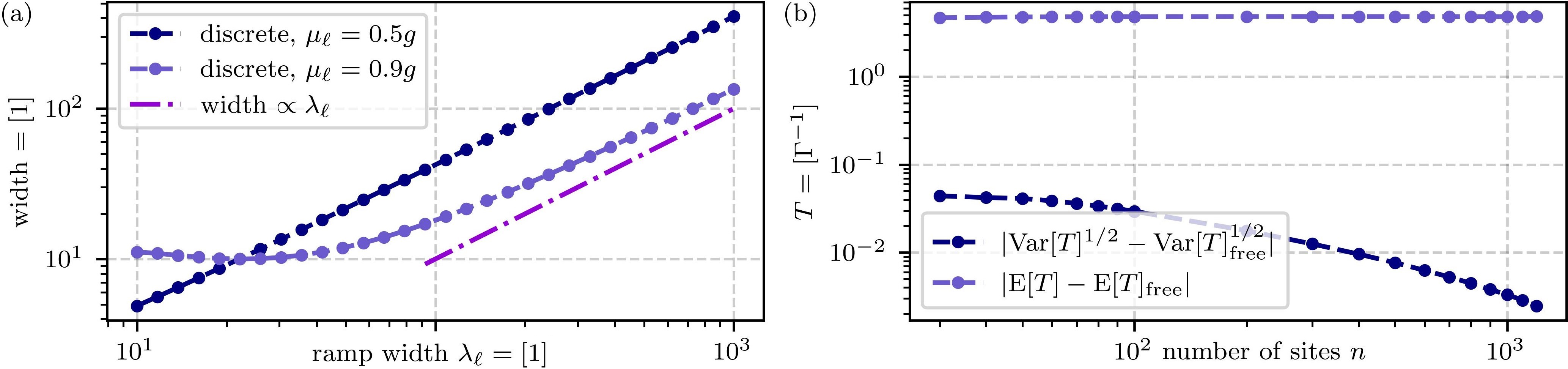}
    \caption{In (a), we show how the width of the wave packet scales with the length $\lambda_\ell$ of the initialization ramp.
    The continuum limit predicts a linear scaling of the width as shown in~\eqref{eq:psi_x_n_f_n} in the limit of large values of $\lambda_\ell.$
    This plot confirms that the scaling holds for large values of $\lambda_\ell$ regardless of the choice of the parameter $\mu_\ell$ which only affects how large $\lambda_\ell$ must be for the continuum limit to dominate also in the discrete case.
    In (b), we show a comparison between the numerically exact tick probability density $P[T=t]$ and the approximation from eq.~\eqref{eq:PTick_loss_approx} where the lossy contribution is neglected.
    We are looking at the difference between the exact average value $\mathrm{E}[T]$ and the one obtained from the approximation $\mathrm{E}[T]_\text{free}$ of the free theory.
    Similarly, we also plot the difference between the standard deviations $\mathrm{Var}[T]^{1/2}$ obtained using the two different methods.
    We find that the error we obtain with this approximation stays constant for the average values and even shrinks for the standard deviation, justifying the approximation made in Sec.~\ref{SM:emission} by neglecting the lossy contributions to the tick PDF.}
    \label{fig:width_scaling}
\end{figure*}

Here, we combine all the previous results to obtain the theoretical result displayed in eq.~\eqref{eq:Nt_num_vs_thy} of the main text.
This result relies on the PDF of the ticks obtained in
Sec.~\ref{SM:calculating_precision},
the preparation by the initial ramp from Sec.~\ref{SM:continuum_description},
the subsequent propagation through the bulk of the ring  from Sec.~\ref{SM:bulk_propagation_dispersion} and eventually the emission or tick event was discussed in Sec.~\ref{SM:emission}.

Our starting point is the PDF in eq.~\eqref{eq:PT=t} and for the following analysis it will be beneficial to decompose the effective Hamiltonian from eq.~\eqref{eq:rho_i_lyapunov} in two parts,
\begin{widetext}
\begin{equation}
    H_{\mathrm{eff}} 
    = 
    \underbrace{\sum_{j=0}^{n-2}\left(-\mu_\ell e^{-j/\lambda_\ell} + g\right)\big(\ketbra{j}{j+1} + \mathrm{h.c.}\big)}_{ = H_{0}}
    +
    \underbrace{\sum_{j=0}^{n-2}\mu_r e^{(j-(n-1))/\lambda_r}\big(\ketbra{j}{j+1} + \mathrm{h.c.}\big) - \frac{i}{2}J^\dagger J}_{ = K }
\end{equation}
where $H_{0}$ describes the dynamics of preparation by the ramp and bulk propagation (see Secs.~\ref{SM:continuum_description} and \ref{SM:bulk_propagation_dispersion}).
We refer to this as the free part of the evolution.
The non-hermitian Hamiltonian $K$ corresponds to dynamics when the wave packet hits the final apodization region and its subsequent conversion to a tick of the clock, we refer to this as the interaction part.
The evolution generated by the effective Hamiltonian $H_{\mathrm{eff}}$ expanded in a Dyson series reads
\begin{equation} \label{eq:SM_Heff_evol}
    e^{-iH_{\mathrm{eff}}t}
    =
    \underbrace{e^{-iH_{0}t}}_{=U(t)}
    +
    \underbrace{e^{-iH_{0}t}
    \sum_{k = 1}^{\infty} {(-i)^k} 
    \int_0^t \mathrm{d}\tau_1 \int_0^{\tau_1}\mathrm{d}\tau_2 \cdots \int_0^{\tau_{k-1}} \mathrm{d}\tau_k\,
    K(\tau_1)\cdots K(\tau_k)}_{=\mathcal V(t)},
\end{equation}
\end{widetext}
where we defined the Hamiltonian in the interaction picture $K(t) = e^{iH_{0}t} K e^{-iH_{0}t}$.
With the decomposition of the effective time evolution in eq.~\eqref{eq:SM_Heff_evol}, the tick probability can also be decomposed into two parts,
\begin{align}
\label{eq:PTick_loss_approx}
    P[T=t] = \underbrace{\Gamma|c_0(t)|^2}_{=p_0(t)} + p_1(t),
\end{align}
where
\begin{align}
    p_1(t) = \Gamma\left(|c_1(t)|^2 + 2\mathfrak{Re}\left[c_0(t)^*c_1(t)\right]\right).
\end{align}
The coefficients $c_0(t)$ and $c_1(t)$ are obtained from the decomposition of $e^{-iH_{\rm eff}t}$ into the free and interacting part, 
\begin{align}
    c_0(t) = \braket{n-1}{U(t)|0},\quad c_1(t) = \braket{n-1}{\mathcal V(t)|0}.
\end{align}
This shows that the tick PDF is made up of two contributions; one exclusively attributed to propagation and the other is the correction coming from apodization as well as emission.

The main result of Sec.~\ref{SM:apodization} is that the couplings $g_j$ as optimized by eq.~\eqref{eq:opt_5} match the right ramp to the tick rate $\Gamma$.
As shown in Fig.~\ref{fig:transmittance}(a,b), the result is that the right ramp transmits the wave packet with close to unit probability without reflection.
Consequently, the tick PDF is, up to small corrections, essentially dominated by the bulk propagation, i.e., the free term $p_0(t)$.
This means that we may neglect the lossy interaction contribution $p_{1}(t)$ without incurring on a significant error for the discussion below.
In Fig.~\ref{fig:width_scaling}(b) a numerical analysis illustrates how much the predicted first and second moments of the tick time differ when using only the free part $p_0(t)$ instead of the full theory $P[T=t]$.

We now use the expression for the propagating wave function in the continuum limit from eq.~\eqref{eq:correction_q3_explicit} and evaluate $c_0(t)$ at time close to the time $n/2g$ where the wave packet has reached the right ramp.
For this, we identify $c_0(t) = \psi_{n-1}(t)$, and we recall that we chose an offset time $t_0$ for obtaining eq.~\eqref{eq:correction_q3_explicit} such that $n\gg2gt_0 \gtrsim \lambda_\ell$, i.e., the wave packet has just left the left ramp but is still far away from the end of the ring.
Then, we can write $c_0(t) = c_0(t_0 + (t-t_0)) \sim c_0(t_0 + t)$ because we are considering times $t\sim n/2g$ much larger than $t_0$.
Looking at small deviations $t = n/2g + \delta t,$ we can write
\begin{equation} \label{eq:c0}
    c_0\left(t\right) 
    \sim 
    \sqrt{\lambda_\ell}\int_{-\infty}^\infty \frac{\mathrm{d}Q}{2\pi} \hat{f}(Q) e^{+iQ \frac{2g\delta t}{\lambda_\ell}} e^{-i\frac{n}{\lambda_\ell^3} \frac{Q^3}{3!}},
\end{equation}
where we dropped the $\delta t \ll n/2g$ term in the $Q^3$ expression.
Note, that the prefactor of the third order term in the exponent  scales like ${n}/{\lambda_\ell^3} Q^3$, which is responsible for the strength of the broadening of the wave packet after traversing the ring.
The skewing of the wave packet thus proliferates with increasing ring size $n$.
However, this expression suggests that, by choosing the appropriate scaling of $\lambda_\ell \sim n^{1/3}$ with the system size, this effect can be mitigated.
This means that the error from the cubic term can be bounded from above by an arbitrarily small constant, and the wave packet thus does not skew as it propagates along the ring.
As discussed in the main text, we find
\begin{equation}
    \lambda_{\ell} \overset{\mathrm{num.}
    }{\sim} 
    n^{0.35}  
    \overset{\mathrm{th.}
    }{\sim} 
    n^{1/3},
\end{equation}
which suggests that the scaling found in the numerical optimization is in good agreement with the necessary scaling predicted by our theory. 
Integrating the expression in eq.~\eqref{eq:c0}, the cubic term can thus be dropped if the asymptotic constant for $n^{1/3}/\lambda_\ell =\varepsilon$ is sufficiently small, and we recover the expression from eq.~\eqref{eq:psi_x_n_f_n} for the absolute square or the tick PDF,
\begin{align}
    P[T = t] \sim \Gamma |c_0(t)|^2 \sim \frac{\Gamma}{n^{1/3}} h\left(\frac{n-2gt}{n^{1/3}}\right)^2,
\end{align}
where
\begin{align}
    h(x) = 
    \int_{-\infty}^\infty \frac{\mathrm{d}Q}{2\pi} \hat{f}(Q) e^{-iQ x -i \varepsilon\frac{Q^3}{3!}} \sim f(x)^2.
\end{align}
Note, here, we have reinstated the absolute time $t$ again. 
The prefactor $n^{-1/3}$ comes from the correct normalization, as also present in eq.~\eqref{eq:psi_x_n_f_n}. 
Due to the strong concentration of $\hat{f}$ we expect that so is $h$ and that $\int\mathrm{d}x\,x h(x) <\infty$ as well as  $\int\mathrm{d}x\,x^2 h(x) <\infty$ are finite. 
Therefore, we find for the first moment of the tick PDF the scaling
\begin{equation}
    \mathrm{E}[T] \sim \frac{\Gamma}{n^{1/3}} \int_{-\infty}^\infty \mathrm{d}t\, t\; h\left(\frac{n-2gt}{n^{1/3}}\right)^2 \sim n,\qquad n\rightarrow\infty.
\end{equation}
For the second moment, i.e., the variance we obtain the scaling
\begin{align}
    \mathrm{Var}[T] &\sim \frac{1}{n^{1/3}} \int_{-\infty}^\infty \mathrm{d}t  \;(t-\mathrm{E}[T])^2 h\left(\frac{n-2gt}{n^{1/3}}\right)^2 \\
    &\sim n^{2/3} \int_{-\infty}^\infty \mathrm{d}x\; x^2 h(x)^2\\
    &\sim n^{2/3},\qquad n\rightarrow\infty.
\end{align}
This scaling is significantly below $\mathrm{Var}[T] \sim n^2$ which we expect from ballistic transport. 
Together with eq.~\eqref{eq:accuracyT} we obtain eq.~\eqref{eq:Nt_num_vs_thy} which is the main result of Sec.~\ref{SE:Ninf}.
In the following section, we give an explanation for how the loss term in eq.~\eqref{eq:PTick_loss_approx} can be neglected due to the boundary matching of the right ramp.

\section{\label{SM:precision_entropy_scaling}Exponential scaling of precision with entropy production}
We now switch back to the regime where the jump process $\overline J$ participates in the evolution.
In the following, we discuss how one can calculate the clock precision as defined in the main text eq.~\eqref{eq:accuracyN(t)},
\begin{align}
    \mathcal N_\Sigma= \lim_{t\rightarrow\infty}\frac{\mathrm{E}[N(t)]}{\mathrm{Var}[N(t)]},
\end{align}
given the generators of the evolution for the ring clock and how $\mathcal N_\infty$ and $\mathcal N_\Sigma$ are related in the infinite entropy production regime (Sec.~\ref{SM:calcualting_precison_reversible}).
Subsequently, we examine what happens in detail if the assumption on the infinite entropy production is dropped. Our goal is to quantify how much $\mathcal N_\Sigma$ and $\mathcal N_\infty$ then differ as a function of the finite entropy production (Sec.~\ref{SM:clock_precision_finite_entropy}).

\subsection{\label{SM:calcualting_precison_reversible}Calculating clock precision in the reversible regime}
Formally, the number of ticks $N(t)$ are defined as an integrated stochastic current $N(t) = \int_0^t\mathrm{d}N(\tau)$.
There, the jumps $\vert n-1\rangle \rightarrow \vert 0\rangle$ ($\vert 0\rangle \rightarrow \vert n-1\rangle$) lead to the incremental increase $\mathrm{d}N(t) = +1$ (decrease $\mathrm{d}N(t) = -1$) respectively.
We can resolve the master equation evolution of the ring clock's state $\rho(t)$ with respect to the number of times the clock has ticked by introducing a free counting field $\chi$~\cite{Schaller2014,Landi2023},
\begin{align}
\label{eq:rho(t,chi)}
    \rho(t,\chi) = \sum_{k\in\mathbb Z} \rho^{(k)}(t)e^{ik\chi}.
\end{align}
The state $\rho^{(k)}(t)$ is the non-normalized clock state given that the net number of ticks at time $t$ is $k.$ The trace $\tr\left[\rho^{(k)}(t)\right]=P[N(t)=k]$ reveals the probability of this event.
To obtain the form $\rho(t,\chi)$ from a dynamical time-evolution, we can introduce the tilted Liouvillian $\mathcal L(\chi)$ of the system,
\begin{align}
\label{eq:L_chi}
    \mathcal L(\chi) = \mathcal L_0 + e^{+i\chi}\mathcal L_+ + e^{-i\chi}\mathcal L_-.
\end{align}
Here, the three parts of the Liouvillian are defined as $\mathcal L_+\,\cdot\,=J\,\cdot\,J^\dagger$ for the jumps counted as positive ticks, and $\mathcal L_-\,\cdot\,=\overline J\,\cdot\,\overline J^\dagger$ for the jumps counted as negative ticks.
Thus, the counting field has a positive sign for the forward jumps $e^{+i\chi}\mathcal L_+$ and a negative sign for the backwards jumps $e^{-i\chi}\mathcal L_-$.
Finally, the term $\mathcal L_0$ generates the evolution conditioned on no jump ocurring, given by,
\begin{align}
    \mathcal L_0\,\cdot\, = -i[H,\,\cdot\,] - \frac{1}{2}\left\{J^\dagger J,\,\cdot\,\right\} - \frac{1}{2}\left\{\overline J^\dagger \overline J,\,\cdot\,\right\}.
\end{align}
The equation $\dot\rho(t,\chi)=\mathcal L(\chi)\rho(t,\chi)$ then generates the time-evolution resulting in the state $\rho(t,\chi)$ of the form as in eq.~\eqref{eq:rho(t,chi)}.
The cumulant generating function $C(\chi,t)$ for $N(t)$ can be obtained using the eigenvalue $\lambda(\chi)$ with the largest real part of the tilted Liouvillian $\mathcal L(\chi)$~\cite{Schaller2014}.
In the long-time limit $t\rightarrow\infty$, it holds that $C(\chi,t) = t\lambda(\chi) + O(1),$ allowing us to determine the asymptotic values for $\mathrm{E}[N(t)]$ and $\mathrm{Var}[N(t)],$
\begin{align}
\label{eq:EN_chi_derivative}
    \lim_{t\rightarrow\infty} \frac{\mathrm{E}[N(t)]}{t} = -i\frac{\rm d}{{\rm d}\chi}\lambda(\chi)\Big|_{\chi=0},
\end{align}
and
\begin{align}
\label{eq:VarN_chi_derivative}
    \lim_{t\rightarrow\infty} \frac{\mathrm{Var}[N(t)]}{t} = -\frac{\mathrm{d}^2}{\mathrm{d}\chi^2}\lambda(\chi)\Big|_{\chi=0},
\end{align}
and therefore the clock precision as in~\eqref{eq:accuracyN(t)}.
These expressions are general and they hold regardless of whether the reverse ticking process generated by $\overline J$ is present or not.

In Sec.~\ref{sec:clock_precision_first_passage_times}, we have calculated the clock precision in terms of the first passage time $T,$ where $\overline J = 0$ was assumed.
This would for example be realized in the limit of infinite entropy production per tick $\Sigma_\mathrm{tick}=\infty$.
If we formally take this limit, the counting variable $N(t)$ falls into the class of renewal-processes~\cite{Cox1962}.
Renewal processes are generalized Poisson processes, where the time between successive events is independently and identically distributed (i.i.d.) according to some waiting time distribution.
For the Poisson process, the waiting time is exponentially distributed, in our case, it is distributed according to $P[T=t]$ as given by the formula~\eqref{eq:PT=t}.
The reason why the time between ticks is i.i.d.\ distributed, is that after every tick, the clock resets to the same initial state is $\rho_0=\ketbra{0}{0}$.
Renewal process have particularly well-behaved properties such that the asymptotic moments of $N(t)$ are related to that of $T$ in the following way~\cite{Cox1962,Silva2023},
\begin{align}
\label{eq:E_N_limit}
    \lim_{t\rightarrow\infty} \frac{\mathrm{E}[N(t)]}{t} = \frac{1}{\mathrm{E}[T]},
\end{align}
and
\begin{align}
\label{eq:Var_N_limit}
    \lim_{t\rightarrow\infty} \frac{\mathrm{Var}[N(t)]}{t} = \frac{\mathrm{Var}[T]}{\mathrm{E}[T]^3}.
\end{align}
In the case of formally infinite entropy production per tick, the clock precision defined with respect to $N(t)$ is therefore the same as the one defined with respect to $T,$ i.e., we have 
\begin{align}
\label{eq:accuracy_N_vs_T}
    \mathcal N_\Sigma = \lim_{t\rightarrow\infty}\frac{\mathrm{E}[N(t)]}{\mathrm{Var}[N(t)]} \rightarrow \mathcal N_\infty =\frac{\mathrm{E}[T]^2}{\mathrm{Var}[T]}.
\end{align}

\subsection{\label{SM:clock_precision_finite_entropy}Clock precision at finite entropy production}
\begin{figure*}[t]
    \centering
    \includegraphics[width=\textwidth]{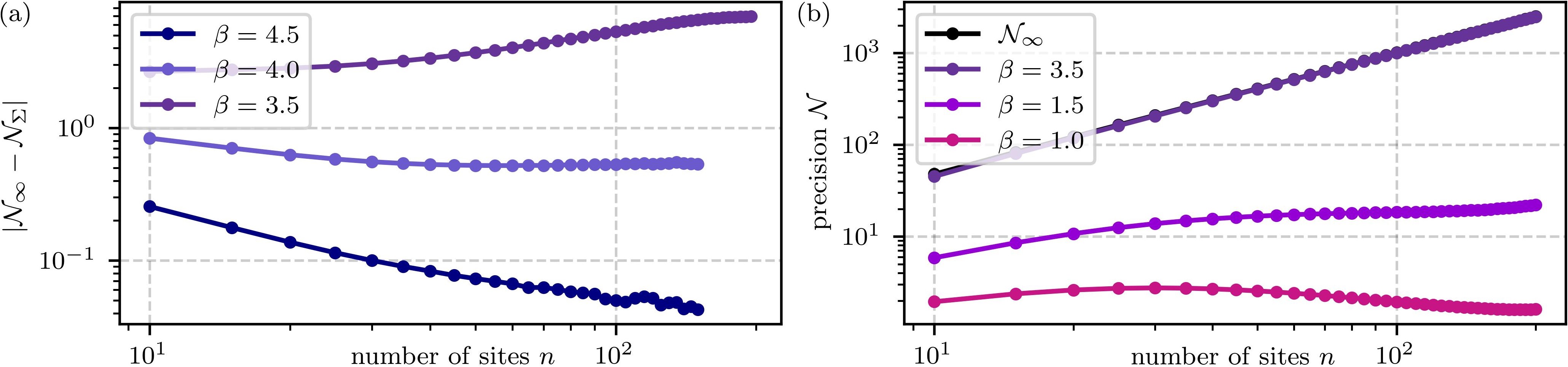}
    \caption{Comparisons between the clock precision $\mathcal N_\Sigma$ in case of finite entropy-production and $\mathcal N_\infty$ in the limit $\delta=0$ are shown.
    (a) We plot the difference $\mathcal N_\infty-\mathcal N_\Sigma$ for different choices of the exponent in $\delta = n^{-\beta}$ in a loglog-scale.
    We see that for all three choices among $\beta \in\{3.5,4.0,4.5\},$ the error of the clock precision is negligible on the scale of $\mathcal N_\Sigma.$
    (b) This is a loglog-plot of the clock precision $\mathcal N_\Sigma$ with a comparison to $\mathcal N_\infty$.
    We look at further choices of $\beta\in\{1.0,1.5,3.5\}$ and we find that if $\beta$ is chosen too small, the perturbation from the reverse tick process destrucively affects the clock precision.
    For $\beta=3.5$, $\mathcal N_\Sigma$ and $\mathcal N_\infty$ overlap on the scale of this plot.}
    \label{fig:accuracy_comparison}
\end{figure*}
Having found a way to determine the moments of $N(t)$ and $T$ and relating them in the case where $\overline J=0,$ the question is raised of how much is the equality~\eqref{eq:accuracy_N_vs_T} violated if $\overline J$ enters as a small but non-zero perturbation.
To deal with this issue we relax the limit $\Sigma_\mathrm{tick}=\infty$ to some finite values of $\Sigma_\mathrm{tick},$
and introduce the perturbative parameter $\delta = e^{-\Sigma_\mathrm{tick}}$ to write $\overline{J} = \sqrt{\delta}J^\dagger.$
We can thus re-express eq.~\eqref{eq:L_chi} as follows,
\begin{align}
    \mathcal L(\chi,\delta) &= -i\left[H,\cdot\right] -\frac{1}{2}\left\{J^\dagger J,\cdot\right\}+ e^{i\chi}J\cdot J^\dagger \nonumber \\
    &\quad + \delta\left(-\frac{1}{2}\left\{J J^\dagger,\cdot\right\} + e^{-i\chi}J^\dagger \cdot J \right). \label{eq:L_chi_delta}
\end{align}
We recover the following two limiting cases:
\begin{itemize}
    \item If $\delta = 0,$ we get eqs.~\eqref{eq:E_N_limit}, \eqref{eq:Var_N_limit} and~\eqref{eq:accuracy_N_vs_T}, with $\mathcal N_\Sigma \rightarrow \mathcal N_\infty.$
    \item If $\delta>0$ is small, the reverse jump $\overline{J}$ enters as a perturbation $\mathcal L(\chi,\delta)= \mathcal L(\chi,0)+O(\delta)$.
    Since $\mathcal N_\Sigma$ is a continuous function of ($\chi$-derivatives) of the dominant eigenvalues of $\mathcal L(\chi,\delta)$, we anticipate that the $\delta$-perturbation of $\mathcal L$ also leads to a $\delta$-perturbation of the precision $\mathcal N_\Sigma=\mathcal N_\infty(1+O(\delta)).$
\end{itemize}
The goal of the following section is to make the second case rigorous in the sense that we determine the prefactors in the $O(\delta)$ perturbation.
We want to end up with a $\delta$ that scales with the ring length as $\delta = n^{-\beta}$ for some value of $\beta>0,$ such that the correction $O(\delta)\rightarrow 0$ vanishes for large ring lengths $n\rightarrow\infty.$
What we not considered explicitly so far: the prefactor in the $O$-notation for the perturbation of the clock precision could also scale with $n.$
For example, if the prefactor were to grow exponentially with $n$, $\delta = n^{-\beta}$ would be insufficient to make the error vanish, and one would have to chose $\delta$ differently.
For reasons detailed in the following analysis, the perturbation is well-behaved and the prefactor only scales polynomially.
The clock precision is therefore perturbed as follows,
\begin{align}
\label{eq:N_NT_expansion}
     \left|\mathcal N_\infty - \mathcal N_\Sigma\right| = O\left(n^{c - \beta}\right),
\end{align}
now including all $n$-dependencies in the $O$-notation.
In Fig.~\ref{fig:accuracy_comparison}(a), we visualize this bound by plotting the absolute difference $|\mathcal N_\infty-\mathcal N_\Sigma|$ and in Fig.~\ref{fig:accuracy_comparison}(b), we show how $\mathcal N_\Sigma$ scales if $\beta$ is not chosen large enough.
The constant $\beta>0$ comes from the choice of $\delta = n^{-\beta}$ and $c>0$ is a constant related to the spectral gap of the ring clock Liouvillian, which we will get to later.
What this tells us is that there exists a choice of $\beta>c,$ where the error between $\mathcal N_\Sigma$ and $\mathcal N_\infty$ becomes negligible for large values of $n.$
We find that $\beta=4$ is a choice that works, as shown in Fig.~\ref{fig:accuracy_comparison}(a).
Using the identification $\delta = e^{-\Sigma_\mathrm{tick}}$ gives us logarithmically growing entropy production per tick $\Sigma_\mathrm{tick} = 4 \log n$.
Combined together with the precision scaling $\mathcal N_\infty \sim n^{1.31}$ that we recover from $\mathcal N_\infty$.
Because their relative error is negligible, we find our desired result,
\begin{align}
     \mathcal N_\Sigma= e^{\Omega(\Sigma_\mathrm{tick})},
\end{align}
clock precision and entropy per tick are exponentially separated.

\paragraph*{Detailed analysis.}
To arrive at the expansion in eq.~\eqref{eq:N_NT_expansion}, we first look at how $\mathrm{E}[N(t)]$ and $\mathrm{Var}[N(t)]$ behave for finite values of $\delta.$
Here, we again assume $\delta$ to be an independent parameter and only in the end we will prescribe a relationship $\delta = n^{-\beta}$.
Formally, we can write the first two moments in terms of a power series in the perturbation $\delta$ as follows by using the results from eq.~\eqref{eq:EN_chi_derivative},
\begin{align}
\label{eq:series_lambda_1k_deltak}
    \lim_{t\rightarrow\infty}\frac{\rm d}{{\rm d}t}\mathrm{E}[N(t)] = \underbrace{-i\lambda_{10}}_{=1/\mathrm{E}[T]} - i\sum_{k=1}^\infty \lambda_{1k}\delta^k,
\end{align}
with the notation $\lambda_{jk} := \partial_\chi^j \partial_\delta^k \lambda(\chi)\big|_{\delta=\chi=0},$ and for later use $\lambda_j := \partial_\chi^j \lambda(\chi)\big|_{\chi=0}.$
For the second moment, we have by using eq.~\eqref{eq:VarN_chi_derivative}
\begin{align}
\label{eq:series_lambda_2k_deltak}
    \lim_{t\rightarrow\infty}\frac{\rm d}{{\rm d}t}\mathrm{Var}[N(t)] = \underbrace{-\lambda_{20}}_{=\mathrm{Var}[T]/\mathrm{E}[T]^3} + \sum_{k=1}^\infty \lambda_{2k} \delta^k.
\end{align}
\begin{figure*}
    \centering
    \includegraphics[width=\textwidth]{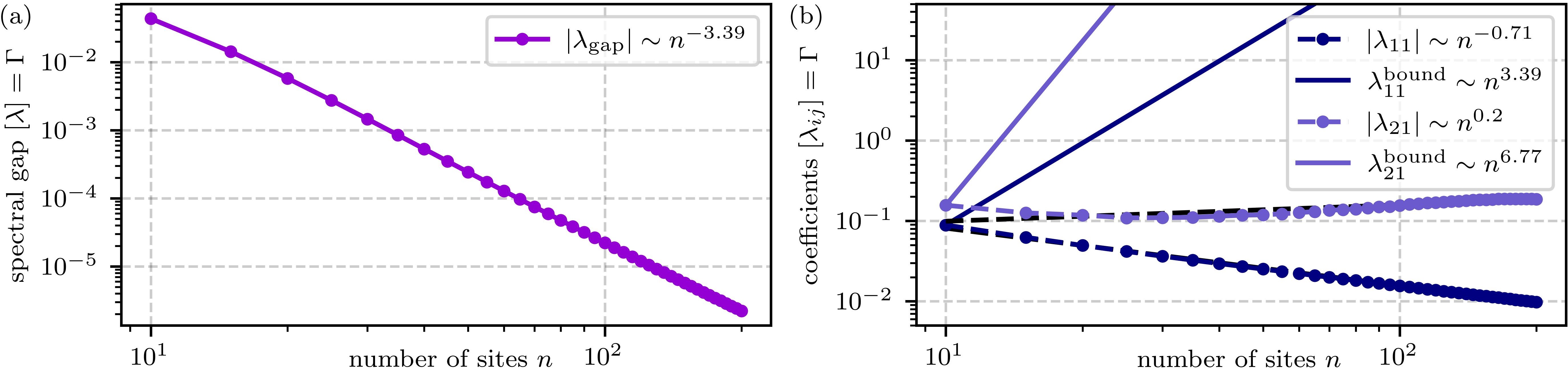}
    \caption{Here we visualize the scaling of the spectral gap and the expansion coefficients of $\mathcal L.$ In (a), we show how the spectral gap of the Liouvillian $\mathcal L$ scales as a function of the number of ring sites $n.$
    The spectral gap is the smallest magnitude non-zero eigenvalue of $\mathcal L$.
    We find that the gap closes only polynomially, which is ultimately the reason why it is sufficient to chose $\delta$ to decay polynomially in $n$ to make the perturbations to the clock precision vanish.
    (b) We show how the first two expansion coefficients of the dominant eigenvalue of $\mathcal L(\chi,\delta))$ scale as a function of $n,$ to visualize the (loose) upper bounds obtained in~\eqref{eq:lambda_1k_lambda_2k_bounds}, using the spectral gap scaling.}
    \label{fig:spectral_gap}
\end{figure*}
The coefficients in the higher order expansion, $\lambda_{jk}$ are also implicitly functions of $n,$ and to ensure that the corrections to eqs.~\eqref{eq:series_lambda_1k_deltak} and~\eqref{eq:series_lambda_2k_deltak}, stemming from finite $\delta,$ are small, these coefficients must not grow too quickly as a function of $n.$
Let us thus examine in more detail these coefficients $\lambda_{jk}$.
They can be determined by looking at the eigenvalue problem $\mathcal L(\chi)\omega(\chi)=\lambda(\chi)\omega(\chi)$ written as a perturbation series in $\delta$ and $\chi.$
The expansion with respect to $\chi$ gives us the derivatives of the dominant eigenvalue $\lambda(\chi)$ which equal the cumulants of $N(t)$ as detailed in eqs.~\eqref{eq:EN_chi_derivative} and~\eqref{eq:VarN_chi_derivative}.
On the other side, the expansion in $\delta$ gives us the two expansion coefficients in~\eqref{eq:series_lambda_1k_deltak} and~\eqref{eq:series_lambda_2k_deltak}.
The eigenvalue problem can be explicitly written as
\begin{align}
    &\left(\sum_{i,j\geq 0} \mathcal L_{ij}\chi^i \delta^j\right) 
    \left(\sum_{i,j\geq 0} \omega_{ij}\chi^i\delta^j\right) \nonumber\\
    &\qquad=\left(\sum_{i,j\geq 0} \lambda_{ij}\chi^i\delta^j\right)
    \left(\sum_{i,j\geq 0} \omega_{ij}\chi^i\delta^j\right).
\end{align}
We can solve this equation iteratively order by order in powers of $\chi$ and $\delta.$
For concreteness, we are interested in expansions in $\chi$ up and including second order, i.e., $i\leq 2,$ and for $\delta,$ we are interested in the whole power series because we want to bound the overall errors entering eqs.~\eqref{eq:series_lambda_1k_deltak} and~\eqref{eq:series_lambda_2k_deltak}.
For the expansion coefficient of $\chi^i \delta^j,$ the equation is given by the following expression,
\begin{align}
    \sum_{k+\ell = i}\sum_{m+n=j} \mathcal L_{km}\omega_{\ell n} = \sum_{k+\ell = i}\sum_{m+n=j} \lambda_{km}\omega_{\ell n},
\end{align}
where we note that $\sum_{k+\ell=i}$ is a shorthand for the sum over the set $\{k,\ell : k+m=i; k,m\in\{0,\dots,i\}\}$ and by definition $\mathcal L_{km}=0$ for indices $m>1.$
The recursive solution works as follows: we know that $\lambda_{00}=0$ is the unique eigenvalue corresponding to the system's steady-state $\omega_{00}$ which we can determine initially.
Then, we may assume that we know all the terms up to $\omega_{i-1,j}$ and $\lambda_{i-1,j}$ or $\omega_{i,j-1},$ and $\lambda_{i,j-1}$.
In the next step, we can determine $\omega_{i,j}$ and $\omega_{i,j}$ as follows. First, we calculate
\begin{align}
\label{eq:lambda_ij}
    \lambda_{ij} = \tr\left[\sum_{k,\ell,m,n\in\Lambda_{ij}}\left(\mathcal L_{km}\omega_{\ell n} -\lambda_{km}\omega_{\ell n}\right)\right],
\end{align}
with the index set $\Lambda_{ij}=\{k,\ell,m,n : k+\ell = i,\, m+n=j,\text{ and } k,\ell\in\{0,\dots,i\},\, m,n\in\{0,\dots,j\}\}\backslash\{(0,i,0,j)\}.$
The terms $km=00,\ell n=ij$ are not present because $\tr[\mathcal L_{00} \,\cdot\,]=0$ and $\lambda_{00}=0.$
Therefore, on the right-hand side of eq.~\eqref{eq:lambda_ij}, only terms of order lower than $ij$ enter, which, by assumption we have already computed.
Next, we can obtain the state $\omega_{ij}$ as follows,
\begin{align}
\label{eq:omega_ij}
    \omega_{ij} = \mathcal L_{00}^+ \left(\sum_{k,\ell,m,n\in\Omega_{ij}}\left(\lambda_{km}\omega_{\ell n} - \mathcal L_{km}\omega_{\ell n}\right)\right),
\end{align}
where $\Omega_{ij}=\Lambda_{ij}\cup \{(0,i,0,j)\}.$
Note that $\lambda_{00}=0$ and thus the expression $\lambda_{00}\omega_{ij}$ does not appear on the right-hand side of eq.~\eqref{eq:omega_ij}.
The operator $\mathcal L_{00}^+$ is the Drazin inverse of $\mathcal L_{00}$ which inverts $\mathcal L_{00}$ except on the subspace of the $0$ eigenvalue~\cite{Landi2023}.
Steps~\eqref{eq:lambda_ij} and~\eqref{eq:omega_ij} together with the base case of the induction, $\lambda_{00}=0$ and $\omega_{00}$ being the steady-state, allow us to compute $\omega_{ij}$ and $\lambda_{ij}$ for arbitrary $i,j.$

From the functional form of $\lambda_{ij}$ and $\omega_{ij}$ we can conclude that $\lambda_{ij}$ is a linear affine function of $(\mathcal L_{00}^+)^{i+j-1},\dots, \mathcal L_{00}^+$.
This is an important observation because if we want to bound how fast $\lambda_{ij}$ grows with $n,$
we need to know how quickly the largest eigenvalue of $\mathcal L_{00}^+$ grows.
The largest eigenvalue of $\mathcal L_{00}^+$, however, is the inverse of the smallest in magnitude and non-zero eigenvalue of $\mathcal L_{00}$, which is also known as the spectral gap $\varepsilon$ of $\mathcal L_{00}.$
If $\varepsilon$ scales as $\varepsilon = \Omega (n^{-\alpha}),$ for some constant $\alpha>0$, the largest eigenvalue of the Drazin inverse $\mathcal L_{00}^+$ grows at most with $n^\alpha.$
Consequently, we can also estimate $\lambda_{ij} = O(n^{(i+j-1)\alpha})$ in the limit of large $n,$ where the contribution from $(\mathcal L_{00}^+)^{i+j-1}$ dominates.
What we were originally interested in is whether the coefficients $\lambda_{1k}$ in eq.~\eqref{eq:series_lambda_1k_deltak} and $\lambda_{2k}$ in eq.~\eqref{eq:series_lambda_2k_deltak} can be bounded by a polynomial in $n$ whose exponent does not grow faster than linearly in $k.$
Given the bounds for the $\lambda_{ij}$ we have just established under the assumption that the spectral gap closes only polynomially, $\varepsilon =\Omega(n^{-\alpha})$, we are guaranteed that
\begin{align}
\label{eq:lambda_1k_lambda_2k_bounds}
    \lambda_{1k} = O(n^{\alpha k})\text{ and } \lambda_{2k} = O(n^{(k+1)\alpha}).
\end{align}
Figure~\ref{fig:spectral_gap}(a) shows that it is indeed the case that the spectral gap only closes polynomially, at least for values of up to $n=200$ that were numerically examined.
The exponent found is $\alpha=3.39$ (rounded to 2 digits).
In Fig.~\ref{fig:spectral_gap}(b), we see that the bounds in~\eqref{eq:lambda_1k_lambda_2k_bounds} are satisfied, but loose.
The underlying reason for the looseness is that in our estimate, we only accounted for the maximum eigenvalue of $\mathcal L_{00}^+$.
However, the other terms in the series may shrink with $n;$ for example, the steady-state populations encoded in $\omega$, become smaller as $n$ grows.
For an upper bound, however, it is sufficient to consider the scaling of the largest values, without taking into account possible terms that improve the scaling in practice.
Coming back to the prescription that $\delta = n^{-\beta},$ it turns out as shown in Fig.~\ref{fig:accuracy_comparison}(a) that a choice of $\beta=4$ is already sufficient to ensure the perturbations from eqs.~\eqref{eq:series_lambda_1k_deltak} and~\eqref{eq:series_lambda_2k_deltak} are negligible for the clock precision.

\end{appendices}
\end{document}